\documentclass[floats,floatfix,showpacs,amssymb,amsmath,prd,twocolumn,superscriptaddress,nofootinbib, aps]{revtex4-2}

\usepackage[utf8]{inputenc}
\usepackage[T1]{fontenc}
\usepackage{hyperref}
\usepackage{bm}
\usepackage{dcolumn}
\usepackage{aas_macros}
\usepackage{amsmath, amssymb}
\usepackage{graphicx}
\usepackage{float}
\usepackage[normalem]{ulem}
\usepackage{soul}
\setstcolor{red}
\hypersetup{colorlinks,bookmarksnumbered,
citecolor=cyan, urlcolor=magenta}

\providecommand{\adsurl}[1]{\href{#1}{ADS}}
\usepackage{natbib}

\newcommand{\mchirp}{\mathcal{M}_c}

\usepackage[usenames, dvipsnames]{xcolor}

\begin{document}
\preprint{APS/123-QED}

\title{Gravitational Wave Detection with Photometric Surveys}

\author{Yijun Wang}
\email{yijunw@caltech.edu}
\affiliation{California Institute of Technology, Pasadena, CA 91125, USA}
\author{Kris Pardo}
\email{kpardo@caltech.edu}
\affiliation{Jet Propulsion Laboratory, California Institute of Technology, Pasadena, CA 91101, USA}
\author{Tzu-Ching Chang}
\email{tzu-ching.chang@jpl.nasa.gov}
\affiliation{Jet Propulsion Laboratory, California Institute of Technology, Pasadena, CA 91101, USA}
\affiliation{California Institute of Technology, Pasadena, CA 91125, USA}
\author{Olivier Dor\'e}
\email{olivier.p.dore@jpl.nasa.gov}
\affiliation{Jet Propulsion Laboratory, California Institute of Technology, Pasadena, CA 91101, USA}
\affiliation{California Institute of Technology, Pasadena, CA 91125, USA}

\date{\today}
\begin{abstract}

Gravitational wave (GW) detections have considerably enriched our understanding of the universe. To date, all GW events from individual sources have been found by interferometer-type detectors. In this paper, we study a GW detection technique based on astrometric solutions from photometric surveys and demonstrate that it offers a highly flexible frequency range that can uniquely complement existing detection methods. From repeated point-source astrometric measurements, periodic GW-induced deflections can be extracted and wave parameters inferred. We emphasize that this method can be applied widely to any photometric surveys relying on relative astrometric measurements, in addition to surveys designed to measure absolute astrometry, such as \textit{Gaia}. We illustrate how high-cadence observations of the galactic bulge, such as offered by the Roman Space Telescope's Exoplanet MicroLensing (EML) survey, have the potential to be a potent GW probe with complementary frequency range to \textit{Gaia}, pulsar timing arrays (PTAs), and the Laser Interferometer Space Antenna (LISA). We calculate that the Roman EML survey is sensitive to GWs with frequencies ranging from $7.7\times10^{-8}~{\rm{Hz}}$ to $5.6\times10^{-4}~\rm{Hz}$, which opens up a unique GW observing window for supermassive black hole binaries and their waveform evolution. While the detection threshold assuming the currently expected performance proves too high for detecting individual GWs in light of the expected supermassive black hole binary population distribution, we show that binaries with chirp mass $\mchirp>10^{8.3}~M_\odot$ out to 100 Mpc can be detected if the telescope is able to achieve an astrometric accuracy of 0.11 mas. To confidently detect binaries with $\mchirp>10^{7}~M_\odot$ out to 50 Mpc, a factor of 100 sensitivity improvement is required. We propose several improvement strategies, including recovering the mean astrometric deflection and increasing astrometric accuracy, number of observed stars, field-of-view size, and observational cadence. We also discuss how other existing and planned photometric surveys could contribute to detecting GWs via astrometry. 
\end{abstract}
\maketitle

\section{Introduction}
\label{bait}

The successful detection of gravitational wave (GW) signals from binary mergers with the Advanced Laser Interferometer Gravitational-wave Observatory (Advanced LIGO) and Virgo collaboration \citep[see e.g.][]{GW190412,GW170817} has spurred great interest in improving detection sensitivity and developing independent detection methods. For GW astronomy, it is crucial that we have access to GWs across as wide a frequency spectrum as possible, since different frequency bands are sensitive to their respective groups of GW sources. A continuous frequency band also allows for observation of the same GW source as it evolves to higher frequencies towards merger, allowing us to extract as much information as possible. 

The Advanced LIGO is sensitive to GWs between $10$ Hz and $7$ kHz \cite{Aasi2015}, ideal for detecting solar-mass binary mergers. The deci-hertz band will be covered by space-based detectors such as TianGo \citep{TianGO2019} and DECIGO \citep{DECIGO2019}, targeting intermediate-mass black hole binaries ($\sim10^2-10^4~M_\odot$ \cite{TianGO2019}). The milli-hertz band will be covered by the space-based Laser Interferometer Space Antenna (LISA) \cite{LISAL3} and TianQin \cite{TianQin2016}. These interferometer-type detectors directly measure the GW-induced change in separation between either suspended or free falling mirrors. In this case, the detector frequency range is limited by noise factors, such as mirror position alignment error, quantum noise and thermal noise \citep[see, e.g.,][]{O3noise2020}. The space-based detectors are sensitive to massive black hole mergers (MBHMs) at high redshifts (e.g. LISA can detect $10^5~M_\odot$ mergers at $z\sim15$ with an SNR of $\sim100$ in the ringdown stage \cite{LISAL3}). Observing MBHMs will be instrumental for modeling black hole evolution history and understanding strong-field gravity features \citep{Thorpe2019}. 

At lower frequencies, interferometer-type detectors are no longer available and there is a frequency gap until the Pulsar Timing Array (PTA) detection method becomes applicable. PTAs measure the Time of Arrival (TOA) of pulses from stable milli-hertz pulsars. Passing GWs modify the pulse frequency, which translates into a timing residual signal. By cross-correlating timing residuals from pairs of pulsars, GW parameters can be extracted \citep{Moore2015,Babak2012,Ellis2012}. The PTA frequency band is limited by mission lifetime as well as the observational cadence. For example, a 5-year survey with an observational frequency of $17~\rm{year}^{-1}$ ($\sim 1/3 ~\rm{week}^{-1}$) \cite{Taylor2016, Verbiest2016} is sensitive to GWs from $6.3\times10^{-9}~\rm{Hz}$ to $5.4\times10^{-7}~\rm{Hz}$. With longer signal integration time and more pulsar pairs, PTAs can detect the supermassive black hole merger background (SMBHMB) as well as individual supermassive black hole binaries (SMBHBs) with chirp mass between $10^4~M_\odot$ and $10^{10}~M_\odot$ \cite{Sesana2008}. Recently, significant evidence for a signal with common amplitude and spectral slope across monitored pulsars was recovered from the 12.5-yr data by the North American Nanohertz Observatory for Gravitational Wave (NANOGrav). However, there was no statistically strong evidence for the quadrupolar spatial correlation expected from a GW background in the General Relativity framework, and it remains to be determined if the observed signal is indeed astrophysical \cite{nanogravGWB}.

We can also detect GWs via astrometry \cite{Book2011,Pyne1996}. Analogous to the theoretical basis for PTAs, passing GWs perturb photon trajectories as they travel from the observed stars to the detector. This perturbation leaves a GW-specific change to the apparent star positions. It is, in principle, possible to extract this change in position from high-precision astrometric data. Similar to the PTA method, the sensitive frequency range depends on both survey lifetime and observational cadence. Accordingly, using astrometric measurements as GW probes is a highly flexible technique since observational frequency is tunable, depending on mission design. Furthermore, we demonstrate in the paper that these GW measurements can be made with relative astrometry and do not require dedicated absolute astrometric missions.

A suitable photometric survey with cadence higher than PTAs can unlock the intermediate frequency band between PTAs and LISA. A survey with such sensitivity range would be able to detect massive black hole binaries from $10^5~M_\odot$ to $10^9~M_\odot$ during inspiral and close to merger. Examples of such sources include the highly eccentric binaries that go near coalescence in the sensitivity range of LISA \cite{Banerjee,Banerjee2}. Detecting these GW sources will add invaluable data for constraining black hole evolution models.

Accessing this frequency range also opens up opportunities for joint analysis of a GW source population using various GW detectors targeting their respective frequency range. Although studies of a variety of massive binary black hole assembly scenarios suggest that most GW sources in the nanohertz band show little frequency drift on the scale of 10 years \cite{sesana_mono1,sesana_mono2}, some sources detectable by LISA start emitting GWs with a potentially detectable signal strength at frequencies lower than the LISA sensitivity limit, such as highly eccentric binaries ejected from stellar clusters due to natal kicks or dynamical processes \cite{Banerjee,Banerjee2}. It is unlikely to observe one GW source migrate across the frequency spectrum, since the inspiral time for sources at the low frequency limit of LISA can be on the order of gigayears \cite{Banerjee2}. However, observing the same population at these different frequencies allows us to piece together ensemble source properties and their evolution.

This astrometric GW detection method in the context of \textit{Gaia} has been studied in detail \cite{Klioner2018,Moore2017}. \textit{Gaia} as a GW probe is sensitive from $10^{-8.5}~{\rm{Hz}}$ to $10^{-6}~{\rm{Hz}}$; at $f>10^{-7.5}$ Hz, \textit{Gaia} will outperform PTA efforts \cite{Moore2017}. In this paper, we discuss how this analysis can in principle be done with astrometric data from any photometric surveys even though they may not provide absolute astrometric measurements as \textit{Gaia} does. As a specific example, we forecast the GW detection sensitivity of the Nancy Grace Roman Space Telescope\footnote{\texttt{https://roman.gsfc.nasa.gov/}}, NASA's next flagship observatory after the James Webb Space Telescope.

The Roman Space Telescope will observe billions of galaxies and thousands of supernovae to probe the time evolution of dark energy and large-scale structure \citep[see, e.g.,][]{Agrawal}. It will perform a micro-lensing survey on the inner Milky Way, as well as high contrast imaging and spectroscopic studies of individual close-by exoplanets \cite{Roman}. For GW detection, its notional Exoplanet MicroLensing (EML) survey is particularly relevant. It is expected to observe $10^8$ stars in 7 fields \cite{Gaudi2019}. It operates in the near-IR with a $\sim0.281\deg^2$ field of view (FoV), with an estimated single-exposure astrometric precision of 1.1 mas \cite{Sanderson2019}. During its nominal lifetime of 5 years, it will survey a total area of $1.97\deg^2$ between Galactic longitudes of $-0.5\deg$ and $1.5\deg$, and Galactic latitudes between $-0.5\deg$ and $-2\deg$. Observational time consists of six 72-day seasons. During each season, the Roman Space Telescope visits the seven fields sequentially and repeats this cycle every 15 minutes. This gives a maximum of $\sim 41,000$ exposures per source, making it ``one of the deepest exposures of the sky ever taken'' \cite{Gaudi2019}.

In this paper, we begin by reviewing the theory for GW-induced astrometric deflections and outlining the general strategy for using photometric surveys as GW probes. We then assess the potential of the Roman EML survey to detect individual binary signals. In Section~\ref{OurNestIsBest}, we discuss directions for performance improvement for photometric surveys similar to the Roman EML survey as GW probes. We then expand to other telescopes and surveys and discuss their potential for astrometrically detecting GWs. 

All of the code used to produce the figures and analysis in this paper is available at: \href{https://github.com/kpardo/estoiles-public}{https://github.com/kpardo/estoiles-public}.

\section{Photometric Surveys as GW Probes} 
\label{AllFishingBoats}

In this section, we first summarize how GW signatures manifest as observable variation in the astrometric solution. We then present estimates of the sensitivity of photometric surveys to GWs as well as their frequency resolution.

\subsection{GW Signature in Astrometry}
In short, a passing GW perturbs the spacetime along the photon trajectory as it travels from the observed star to the detector. This perturbation causes a shift in the stellar apparent position from its true position. Theoretical details are derived in \cite{Pyne1996} in the distant source limit and later generalized in \cite{Book2011}. Here we present a brief summary, closely following steps in \cite{Book2011}. 

We start with the model where the GW source and observer are stationary in Minkowski spacetime and the GW is a linear perturbation to flat spacetime. Throughout this paper, we use Greek alphabet to denote components of $4$-vectors and Latin alphabet to denote the spatial dimensions. Indices that appear both as upper and lower indices imply summation over all dimensions. We also adopt the transverse-traceless gauge. Under this gauge condition, components of the perturbation tensor, $h_{\mu\nu}$, can be non-zero only when both indices are spatial, and the tensor trace is 0, i.e.: $$h_{0\mu}=0,~ h^\mu_\mu=0 \; .$$
The metric can then be written as:
\begin{equation}
    ds^2=-dt^2+(\delta_{ij}+h_{ij})dx^idx^j \; .
\end{equation}
We can write the photon trajectory as: 
\begin{equation}
    x^\alpha (\lambda) = x_{(0)}^\alpha (\lambda) +x_{(1)}^\alpha (\lambda) \; ,
\end{equation}
where subscript $(0)$ indicates quantities in unperturbed spacetime, and subscript $(1)$ indicates first order corrections. $\lambda$ is the associated affine parameter. 
We calculate the Christoffel symbols in this metric and write the geodesic equation as: 
\begin{eqnarray}
    \frac{d^2x_{(1)}^0}{d\lambda^2}&=&-\frac{\omega_0^2}{2}n^in^jh_{ij,0} \\
    \frac{d^2x_{(1)}^k}{d\lambda^2}&=&-\frac{\omega_0^2}{2}\Big[-2n^ih_{ki,0} \nonumber \\
    && +n^in^j\big(h_{ki,j}+h_{kj,i}-h_{ij,k}\big)\Big] \; ,
\end{eqnarray}
where $\omega_0$ is the photon frequency without GW perturbation. Integrating the geodesic equation with respect to $\lambda$ gives the photon trajectory and $4$-momentum. 

We then compute the GW perturbation in the observer frame. We first construct an orthonormal tetrad, $e_{\hat{\alpha}}$ where $e_{\hat{0}}=\vec{u}$ and $\vec{u}$ is the observer's $4$-velocity. We also require this tetrad to be parallel-transported along the observer worldline. Imposing the parallel-transport equation and the metric, we can express the observer tetrad in terms of the GW and the unperturbed basis vectors. The observed photon $4$-momentum, $k^{\hat{\alpha}}$, can be found via a coordinate transformation, and its spatial part gives $n^{\hat{i}}$. Assuming small deflections, $dn^{\hat{i}} = n^{\hat{i}}-n_{(0)}^{\hat{i}}$. 

It is oftentimes useful to assume monochromatic plane-wave GWs and a distant source, in which case the observed star is many GW wavelengths away from the observer. In the plane-wave model, the integral along geodesics can be done analytically, resulting in some geometrical constant factors and a phase in the form of $e^{-i2\pi f\omega_0(1+\mathbf{p}\cdot\mathbf{n})\lambda_s}$, where $2\pi f$ and $\omega_0$ are frequencies of the GW and the photon, respectively. $\mathbf{p}$ is the GW propagation direction and $\mathbf{n}$ points towards the observed star. In the distant source limit (i.e., $\omega_0\lambda_s \gg c/2\pi f$), prefactors to the integral become negligibly small and we may ignore this term. Consequently, the leading order of the signal depends only on the GW amplitude at the observer. $dn^{\hat{i}}$ is thus much simplified and becomes \cite{Book2011,Pyne1996}: 
\begin{equation}
\begin{split}
    \label{eqn:dn}
    dn^{\hat{i}}(t,\mathbf{n}) = \frac{n^i+p^i}{2\left(1+\mathbf{p}\cdot\mathbf{n}\right)}&h_{jk}(t,\mathbf{0})n^jn^k\\&-\frac{1}{2}h^{ij}(t,\mathbf{0})n_j \; ,
\end{split}
\end{equation}
\begin{equation}
    h_{ij}(t,\mathbf{x}) ={\rm{Re}}\left[\mathcal{H}_{ij}e^{-i2\pi f \left(t-\mathbf{p}\cdot\mathbf{x}\right)}\right] \; ,
\end{equation}
where $\mathcal{H}_{ij}$ denotes the plane wave amplitude tensor. The distant source approximation is also adopted in PTA analyses, where the integral from the pulsar to the observer is reduced to consideration about the two end points only (see, e.g.,\cite{Anholm2009}). For PTA analyses, an additional reason to drop the GW term at the pulsar is that such a signal would be uncorrelated between different pulsars, whereas the GW perturbation at the detector is shared. When we consider the correlation between timing residuals, these pulsar perturbation terms can thus be treated as random noise \citep{Moore2015}. In Section~\ref{OurFishingBoat} we discuss the validity of this assumption in our work.

For small astrometric deflections, it suffices to consider the leading order of $\mathcal{H}_{ij}$ \citep[see e.g.][]{Blanchet2014}, 
\begin{equation}
    \mathcal{H}_{ij} = AH_{ij}(\mathbf{p})
\end{equation}
\begin{equation}
\label{eqn:h0}
    A_{(0)} = \frac{2G^{5/3}}{c^4}(\pi f)^{2/3}\frac{\mchirp^{5/3}}{D_L} \sim f^{2/3}M_s^{5/3}  \;.
\end{equation}
$H_{ij}(\mathbf{p})$ is the polarization tensor for GWs propagating along $\mathbf{p}$. $A_{(0)}$ is the leading term of the GW amplitude, $A$, which depends on the source frame GW frequency, $f$, the chirp mass, $\mchirp$, and the luminosity distance, $D_L$. $\mchirp$ is defined as $m\left(q/(1+q)^2\right)^{3/5}$, where $m$ is the total mass of the binary and $q$ is the mass ratio, $m_1/m_2$, assuming $m_1$ is the smaller mass. For GW sources not at cosmological distances (i.e., redshift $z\ll 1$), we may ignore the cosmological redshift to the wave frequency. Throughout this paper, we always assume such close-by sources and we do not differentiate between source frame and observer frame GW frequency. Our threshold GW source estimates validate this assumption. To this order, we note that scaling $\mchirp$ by an arbitrary factor $\kappa$ is completely degenerate with scaling $D_L$ by $\kappa^{5/3}$. Therefore, it is convenient to define a scaled mass $M_s \equiv \mchirp/D_L^{3/5}$, which represents all sources that give the same leading order GW signal, at a fixed frequency. 

\begin{figure*}[!htb]
    \centering
    \includegraphics[width=.8\textwidth]{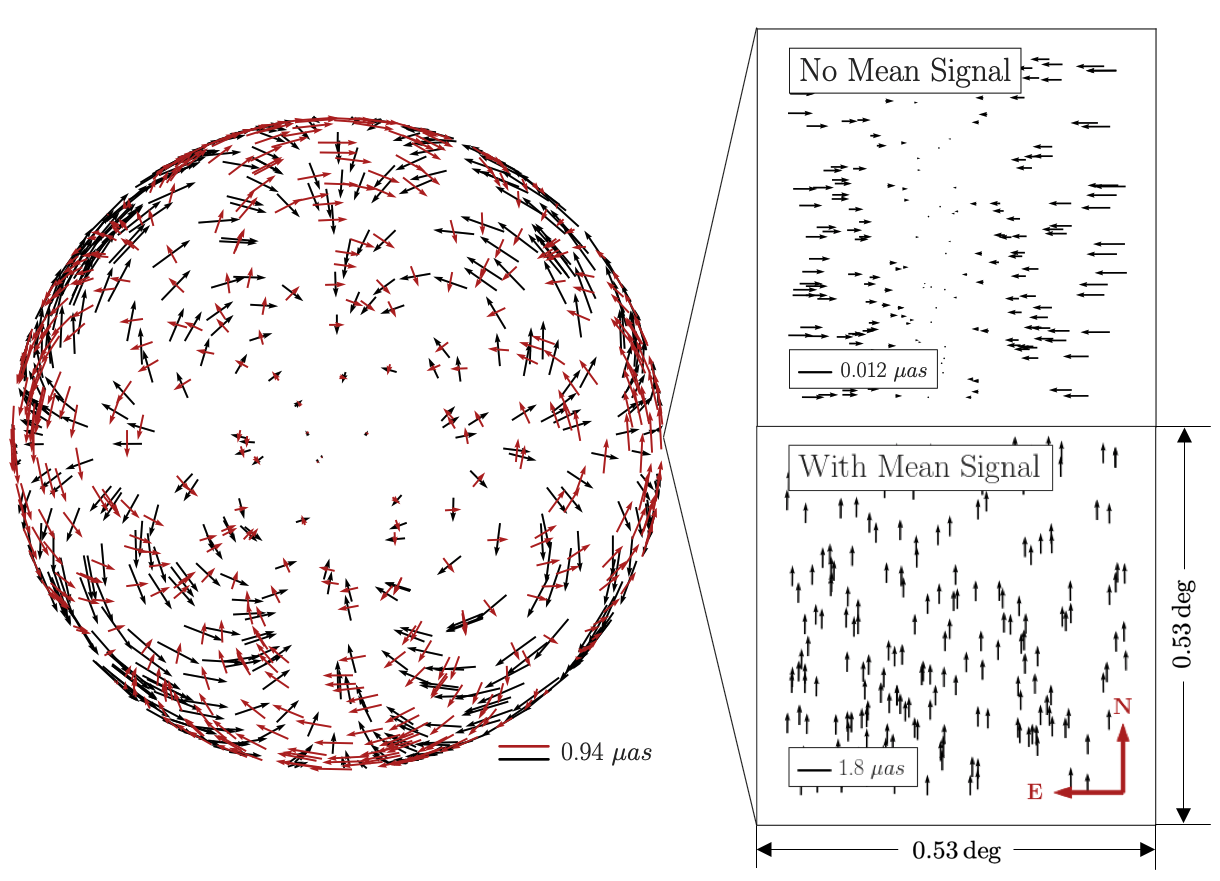}
    \caption{\small{Illustration of the expected stellar astrometric deflections. \textit{Left:} Orthographically projected $dn$ for a subset of stars observed by \textit{Gaia} in the northern hemisphere onto the galactic plane \citep[inspired by a very similar plot in][]{Moore2017}. The North Galactic pole is at the center which is also the position of the GW source. Black arrows correspond to the real part of the waveform at GW phase $\phi=0$ (plus polarization), and the red arrows correspond to that at $\phi=\pi/4$ (cross polarization). The source is a $10^9~M_{\odot}$ equal-mass binary black hole at $1~\rm{Mpc}$ at $(l=90\deg,b=90\deg)$ in galactic coordinates, emitting GWs at $10^{-6}~\rm{Hz}$. This inclination angle is set to $i=0$ (i.e. face-on) and the polarization angle is $\psi=0$. \textit{Right:} Deflections within the Roman Space Telescope's FOV during the EML survey. The lower panel shows the total deflection, and the upper panel shows the deflection after subtracting the mean, since the mean is expected to be absorbed in the pointing reconstruction; for further discussion see Section~\ref{OurNestIsBest}. Star coordinates are selected from the Gaia Data Release 2 catalog, with brightness $0<G<9$ \cite{Gaia2016, GaiaDR2_2018}. Density of stars reflects only a subset of the true stellar density in the catalog.}}
    \label{fig:dnFoV}
\end{figure*}

In Figure~\ref{fig:dnFoV} we reproduce Figure 1 in \cite{Moore2017} and illustrate the astrometric deflection pattern for a field of stars in the northern hemisphere in Galactic coordinates, due to a face-on GW source at zenith. It is clear that the deflection magnitude is largest on the Galactic plane. Deflections induced by the plus and cross polarizations are orthogonal, and the quadrupolar pattern is clear. The right panels show the astrometric deflection in a square Field of View (FoV), assuming the telescope is in the Galactic plane and points directly to the Galactic center. This FoV model has roughly the same area as the true FoV of the Roman Space Telescope but differs in shape. We adopt it nonetheless in our analysis for simplicity. 

The bottom panel shows the total deflection pattern while the upper panel shows the deflection pattern after subtracting the mean deflection. This is expected to be the actual observed signal, as the pointing reconstruction strategy of the Roman Space Telescope will likely absorb deflections uniform across the FoV. A measure of the magnitude of the mean-subtracted deflections is the divergence of $dn$ integrated across the FoV, since the mean deflection field has zero divergence. For the particular GW source position and telescope pointing in Figure~\ref{fig:dnFoV}, we compute the integrated divergence of the astrometric deflection in Equation~\ref{eqn:dn} to be $Al^2_{\rm{FoV}}$ assuming small FoV side length $l_{\rm{FoV}}$, where $A$ is the GW amplitude. From the top right panel in Figure~\ref{fig:dnFoV} and the divergence theorem, the integrated divergence is proportional to $l_{\rm{ FoV}}\langle |dn_{\rm{ms}}|\rangle$, where $\langle |dn_{\rm{ms}}|\rangle$ is the average magnitude of the mean-subtracted deflections. This scaling relation is confirmed numerically using various $l_{\rm FoV}$. For any small-FoV, relative-astrometry telescope, we may estimate the observable deflection signal by 
\begin{equation}
    \langle |dn_{\rm ms}| \rangle \approx \frac{l_{\rm FoV}}{l_{\rm{FoV,RST}}}\langle |dn_{\rm ms,RST}| \rangle\;,
\end{equation}
where the $\rm RST$ subscript denotes parameter values in the case of the Roman Space Telescope. For example, in a FoV similar to the Hubble Space Telescope ($l_{\rm{FoV}}\approx 2.4~\rm{arcmin}$ \cite{Hubble}), the mean magnitude of the mean-subtracted deflections is only $7.4\%$ of that in a FoV similar to the Roman Space Telescope, which has a $l_{\rm{FoV}}\approx 32~\rm{arcmin}$. For further discussions see Section~\ref{OurNestIsBest}.

\begin{figure}
    \centering
    \includegraphics[width=.5\textwidth]{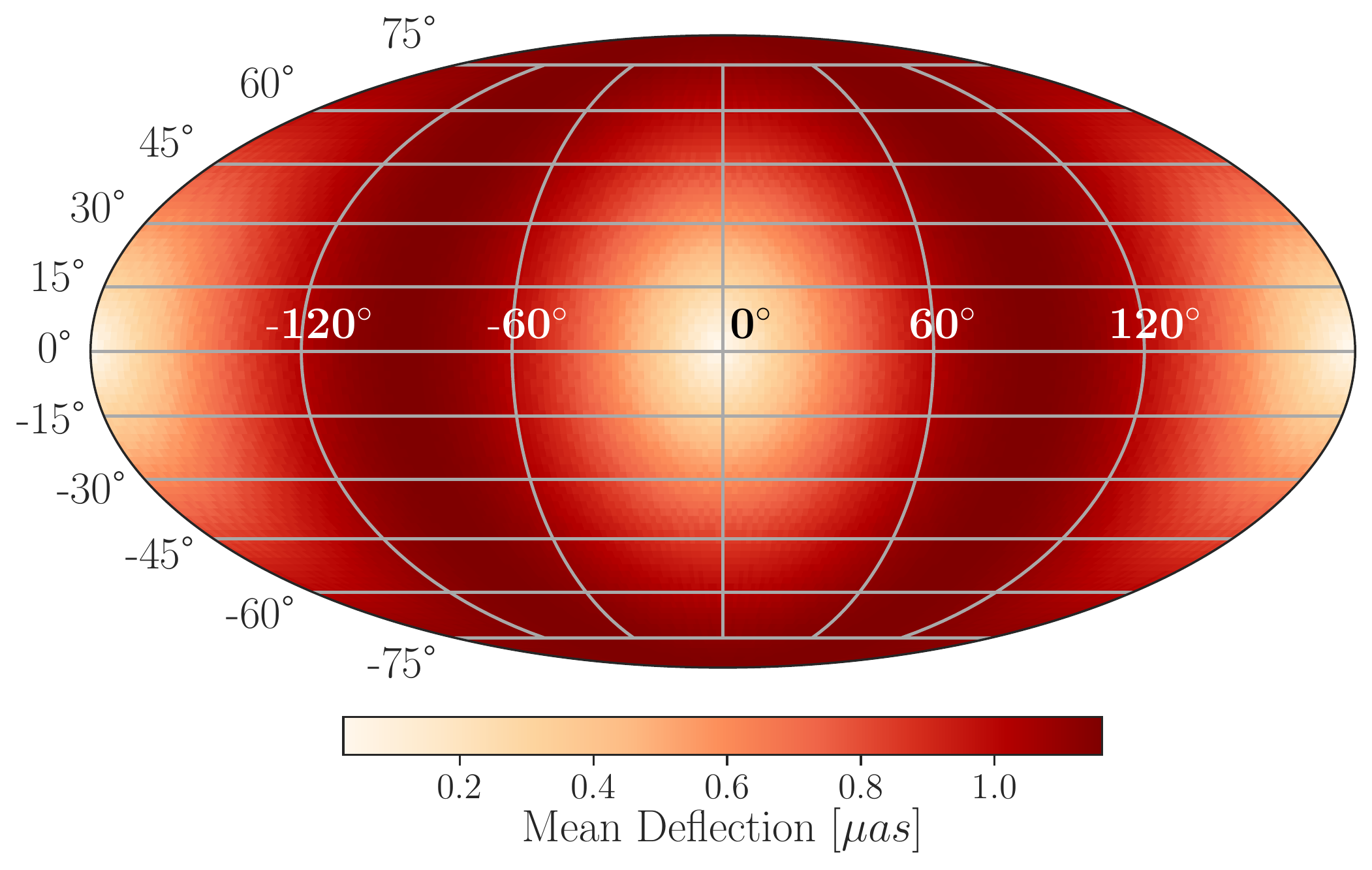}
    \caption{\small{Mean maximal deflection due to GW sources at different Galactic coordinates. The FoV is fixed to point towards the Galactic center, i.e. $l=0\deg,~b=0\deg$. The magnitude is calculated as that of the orthographic projection of $dn$ in the FoV, averaged over all observed stars. The maximum occurs when the source position vector is perpendicular to the star position vectors, which is consistent with Equation~\ref{eqn:dn}. Aside from its coordinates, the GW source at each position has the same properties as that in Figure~\ref{fig:dnFoV}. In combination with Figure~\ref{fig:dnFoV}, we observe that the quadrupolar
    deflection pattern does not show up when we consider signal magnitude only.}}
    \label{fig:heat_map}
\end{figure}

The magnitude of the astrometric deflection as a function of the GW source position on the sky is shown in Figure~\ref{fig:heat_map}. We assume the telescope FoV points to the Galactic center. Properties of this GW source are the same as in Figure~\ref{fig:dnFoV}. For illustration purpose, we fix the polarization angle to be 0. The mean deflection is averaged over 1000 randomly distributed stars within the FoV (the number of stars is not representative of the actual stellar density; it is picked for clear visualization) The deflection is maximal when the GW source position is orthogonal to observed star positions, which is consistent with Equation~\ref{eqn:dn} and Figure~\ref{fig:dnFoV}. 

\subsection{Sensitivity Curve Estimate}
For single exposures, the astrometric accuracy, $\Delta \theta$, is determined by pixel size and pixel placement error \cite{Sanderson2019}. Typically, astrometric deflections due to GWs are small compared to any realistic single-exposure resolution values, therefore they cannot be resolved from isolated measurements of individual stars. This limit, however, can be statistically improved by considering repeated observation of a vast collection of stars. 

Firstly, within each exposure, we consider the correlated astrometric deflection between $N_s$ stars, which improves the astrometric resolution by $\sqrt{N_s}$. Secondly, if the same stars are measured for $N_m$ times throughout the survey, the measurement error is reduced by $\sqrt{N_m}$, which transforms the single-measurement resolution to the end-of-survey resolution. Assuming the same observational cadence throughout the survey, $N_m = T_{\rm{obs}}/\Delta t$, where $T_{\rm{obs}}$ is the total observation time and $1/\Delta t$ is the observational cadence. The minimum detectable GW amplitude is then 
\begin{equation}
\label{Eqn:hc}
    h =\frac{\Delta \theta}{\sqrt{N_sN_m}}\;.
\end{equation}

These two statistical improvements are subject to survey-specific constraints. In Section~\ref{OurFishingBoat}, we provide further discussion on this limit in the context of the Roman EML survey. Throughout our analysis, we assume that, within certain limits, high frequency oscillations are sampled equally well as low frequency ones, i.e. without including discrete sampling effects. In reality, the discrete telescope schedule to visit a sky patch and mission duty cycle impose an upper limit on the maximum number of observable cycles, i.e., the GW frequency, before the deflections become poorly sampled. 

We emphasize that this estimated sensitivity is valid only when entire deflection signals are observable. For telescopes taking relative astrometric measurements, such as the Roman Space Telescope, the observable is, in fact, a small fraction of this total signal, which lowers the sensitivity. See Section~\ref{OurFishingBoat} for detailed discussions.

\subsection{Frequency Resolution}

In this subsection, we outline how to calculate the GW frequency resolution of a photometric survey. From the instrument perspective, the frequency resolution is determined by the exposure timing accuracy, $\sigma_t$. To calculate how $\sigma_t$ translates into end-of-mission frequency resolution, $\Delta f$, we model the GW phase, or equivalently, the phase of the astrometric deflection signal, as a quadratic function of the time of exposures, 
\begin{equation} 
\begin{split}
    \phi 
    &=\phi_0+
    \frac{d\phi}{dt}t+\frac{1}{2}\frac{d^2\phi}{dt^2}t^2\\
    &=  \phi_0+2\pi f t+\pi \dot{f}t^2\;.
\end{split}
\end{equation}

Using Fisher information theory, the uncertainty of the coefficients are
\begin{equation}
\begin{split}
    \sigma(f) &\approx \frac{\sqrt{12}}{T_{\rm{obs}}}\frac{f\sigma_t}{\sqrt{N_m}}\\
    \sigma(\dot{f}) & \approx \frac{\sigma(f)}{T_{\rm{obs}}} \sim \frac{f\sigma_t}{T^2_{\rm{obs}}\sqrt{N_m}}\;,
\end{split}
\end{equation}
where, again, $T_{\rm{obs}}$ and $N_m$ is the total number of exposures.

The campaign frequency sensitivity is then estimated by 
\begin{equation}
    \Delta f =\sigma_{\dot{f}}T_{\rm{obs}}\;.
\end{equation}

To determine whether this resolution is sufficient to capture frequency change of GWs within the detector frequency band, we compare $\Delta f$ with the frequency evolution of observable sources. The intrinsic inspiral binary frequency is given by \citep[see, e.g.,][]{Kidder}
\begin{equation}
    f \sim \frac{1}{m\pi}\left(\frac{1}{4\eta^{1/4}}\left(1+\eta_1\Theta^{-1/4}\right)\right)^{3/2}\;,
\label{eqn:f_ins}
\end{equation}
where $\eta\equiv q/(1+q)^2$ and $m$ and $q$ are the total mass of the binary (in natural units where $G=c=1$) and the mass ratio, as defined before. $\eta_1$ is defined as $743/4032+11/48\eta$ and $\Theta$ as $\eta\left(t_c-t\right)/5m$. Time to coalesce $t_c$ is $5m/\left[\eta\left(8\pi mf\right)^{8/3}\right]$. For systems that remain in inspiral stage throughout survey time, we consider the difference of GW frequencies evaluated at the beginning and the end of the survey. For systems that merger within observational time, we take the final frequency to be the Innermost Stable Circular Orbit frequency, $f_{\rm{ISCO}}$, beyond which the systems quickly coalesce and Equation~\ref{eqn:f_ins} no longer captures the actual frequency.

This characteristic frequency progression could help distinguishing the GW signal from other noise factors and provide additional information for GW source parameter estimation.

\section{Detecting GWs with the Roman Space Telescope} 
\label{OurFishingBoat}

In this section, we explore the potential of the Roman EML survey as a GW probe. We first discuss its sensitivity frequency range following the method outlined in Section~\ref{AllFishingBoats}. Since the procedure is general, we apply a parallel analysis on \textit{Gaia} for comparison. We then describe a method to extract GW signals from photometric data via Bayesian inference. We apply this technique to the Roman Space Telescope and calculate its sensitivity curve. 

\subsection{Roman EML Survey Sensitivity Curve}

Similarly to PTAs, the GW frequency band of photometric surveys is constrained by the observation time span and cadence. At the low frequency limit, the GW period should not be longer than the observation time. Signals with longer period are close to being linear over the observational window and thus are likely to be absorbed as telescope motion or proper motion in the astrometric solution. At the high frequency limit, the GW period should not be shorter than twice the observational cadence to satisfy the Nyquist-Shannon sampling condition.

In addition, the low frequency limit is subject to more detailed and survey-specific modifications. Firstly, the low frequency limit where $f_{\rm{min}}\approx 1/T_{\rm{obs}}$ can be technically relaxed to $f_{\rm{min}}\approx 1/2T_{\rm{obs}}$, since the former limit still produces an oscillatory signal that cannot be fully absorbed by any linear proper motion model \cite{Klioner2018}. However, we note that this only leads to a factor of two difference, and we ignore this factor when estimating the frequency range. Furthermore, the general guideline works best for uniform sampling, whereas actual surveys may have significant periods of downtime between observational windows. In this case, detecting low frequency GWs requires precisely piecing together high-cadence observation seasons that may be quite separated in time. Deflection change within each season is only a fraction of the total amplitude, and may well be approximated by linear proper motion. Considering the magnitude of the signal and uncertainties from long season-separation, this wave reconstruction process will likely introduce large errors that render further data analysis unfeasible. For a conservative limit, $f_{\rm{min}}$ is $1/2T_s$ where $T_s$ is the length of one observational season. 

Specifically for the Roman EML survey, we assume a 15-minute cadence with six 72-day observational seasons spread out over the nominal 5-year mission time. The previous constraints then give a conservative frequency range as:
$$7.7\times10^{-8}~{\rm{Hz}}<\Omega_{\rm{Roman}}<5.6\times10^{-4}~\rm{Hz} \; .$$

We also assume a single-exposure astrometric accuracy of $\Delta \theta \sim 1.1~\rm{mas}$, estimated for $H_{AB}=21.6$ stars,  \citep{Sanderson2019} and a total of $N_s\sim10^8$ stars with $W145_{AB}<23$ \citep{Gaudi2019}. We note that all GW signals within the Roman EML survey frequency range have wavelengths smaller than $\sim0.1~\rm{pc}$, which is much smaller than the distance to any stars Roman Space Telescope observes. Therefore we may safely use the distant source limit described in Section~\ref{AllFishingBoats}. 

We now calculate the frequency resolution, following the procedure in Section~\ref{AllFishingBoats}. Taking a conservative timing accuracy of 1 s and $T_{\rm{obs}}=6\times 72~\rm{days}$ (i.e., assuming all seasons happen consecutively), we estimate $\Delta f$ to be $\sim 10^{-14}$ Hz and $\sim10^{-12}$ Hz for signals at the lower and upper frequency band limit, respectively. For light systems ($\log_{10}\mchirp[M_\odot]=5.7$), intrinsic frequency change of GWs during the inspiral ranges from $10^{-11}~\rm{Hz}$ to $10^{-2}~\rm{Hz}$, depending on its frequency at the start of the observation. If such a system is initially observed to emit GW at $\sim 6\times10^{-5}$ Hz or higher, it coalesces within $T_{\rm{obs}}$. For heavy systems ($\log_{10}\mchirp[M_\odot]=9.7$), GW frequency change ranges from $10^{-8}~\rm{Hz}$ to $10^{-6}~\rm{Hz}$. Such heavy systems coalesce within the observational time window if they emit GW at $\sim 2\times10^{-7}$ Hz at the start of the mission. In all cases, the Roman EML survey will be sensitive to the frequency evolution of detected GWs. We note that it should increase the sensitivity of the Roman EML survey to GWs; however, a full analysis of this effect is outside the scope of this work. 

For \textit{Gaia}, assuming 70 evenly-spaced visits of the same stars, uniformly spread out over the nominal 5-year mission time \cite{Gaia2016}, the frequency range is: $$6.3\times10^{-9}~{\rm{Hz}}<\Omega_{\rm{\textit{Gaia}}}<4.5\times10^{-7}~\rm{Hz} \; . $$ This range differs from \cite{Klioner2018} at the upper limit, since they used the \textit{Gaia} rotational period of $\sim6\rm{h}$ as the cadence. Since we are interested in the average sensitivity applicable to the majority of the observed stars, we adopt the more conservative cadence of $70/5$-year. We adopt $\Delta \theta\sim 0.7$ mas, which is the parallax uncertainty for $G\sim20$ stars in \textit{Gaia} Data Release 2 \footnote{In the recent \textit{Gaia} Early Data Release 3, the standard uncertainty in declination at epoch J2016.0 for magnitude $G=20$ stars is 0.382 mas \cite{eDR3}. The performance estimate we give here will not be drastically different.}\cite{Luri2018}. This magnitude threshold value is picked for convenient comparison with the Roman EML survey, where relatively fainter stars could also be observed in the near IR. We assume $N_s \sim 10^{9}$ \cite{GaiaDR2_2018}.

\begin{figure}
    \centering
    \includegraphics[width=\columnwidth]{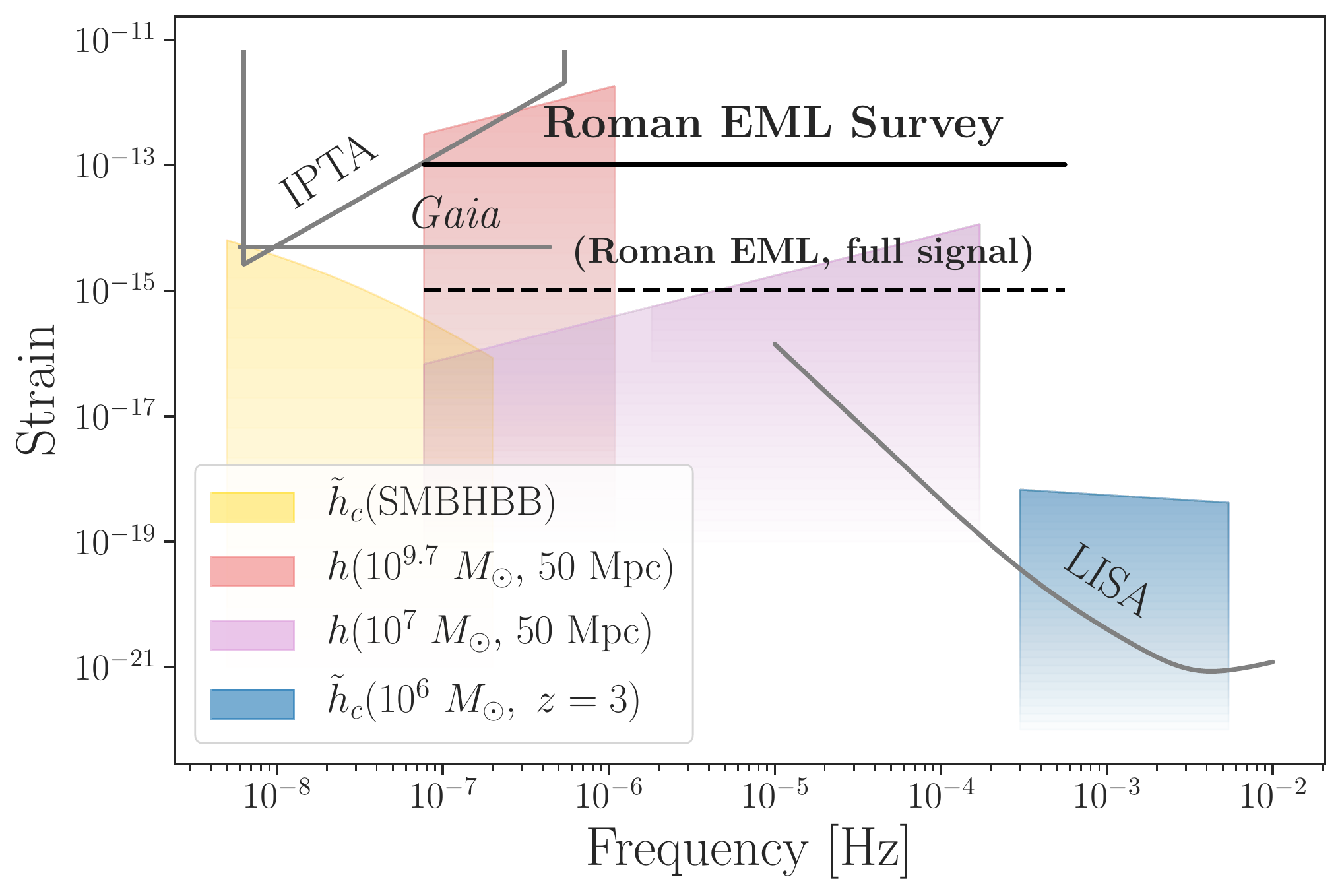}
    \caption{\small{Strain sensitivity of various GW detectors and corresponding example signals. Note that the sensitivity for different detectors is not represented by the same quantity, in anticipation of the signal source types. The sensitivity for LISA and IPTA is represented by the dimensionless characteristic noise strain from the detector noise power spectral density, given by $\tilde{h}_n(f)=\sqrt{fS_n(f)}$. In the frequency range of IPTA, the yellow block shows the expected $\tilde{h}_c$ of the supermassive black hole binary background (SMBHBB). In the frequency range of LISA, the blue block shows the characteristic strain, $\tilde{h}_c(f)\equiv 2f\tilde{h}(f)$, of a fast-evolving $10^6~M_\odot$ binary at $D_L=25~$ Gpc. The sensitivity of \textit{Gaia} and Roman EML Survey is represented by the detectable instantaneous (time-domain) strain, $h$, of monochromatic GWs, assuming end-of-survey performance. For Roman EML Survey, the solid line shows the sensitivity under signal mean subtraction, and the dashed line shows the sensitivity if full astrometric deflections are detectable. The red and violet block show the time-domain wave amplitude of monochromatic GWs with chirp masses $10^{9.7}~M_\odot$ and $10^7~M_\odot$. The shown frequency range of these GWs are limited by the mission lifetime of Roman EML survey and $f_{\rm ISCO}$ at this chirp mass. Note that the example signals of Roman EML and LISA are illustrative; they do not reflect the GW source population expected from binary black hole formation theory. As is shown, the frequency band from roughly $5\times10^{-7}$ Hz to $1\times10^{-5}$ Hz is uniquely accessed by Roman EML survey.}}
    \label{fig:hc_cam}
\end{figure}

Applying Equation~\ref{Eqn:hc} to the Roman EML survey and \textit{Gaia}, we show their strain sensitivity in Figure~\ref{fig:hc_cam}. For the Roman EML survey, the result of Equation~\ref{Eqn:hc} is shown as the dashed line; it is the sensitivity if the Roman EML survey can capture the mean astrometric deflection signal. The sensitivity with the mean-subtracted signals is shown as the black solid line, assuming an average sensitivity decrease of 100. We reiterate that the Roman Space Telescope takes relative astrometric measurements, recording only the relative positions of objects with each other and across the exposures with its nominal astrometric resolution. The absolute positions will be determined by the guiding stars and the telescope pointing with a larger uncertainty. Consequently, it is this mean-subtracted/relative measurements that constitute the data for extracting GW signals. See Figure~\ref{fig:thresh} for an illustration of this scaling. Sensitivity curves for the International Pulsar Timing Array (IPTA) \citep[see][]{Taylor2016, Verbiest2016,Thrane2013} and LISA \cite{Robson2019} are shown for comparison. It is important to note that, due to the targeted signal type, the sensitivity is represented by different quantities for each detector. For IPTA and LISA, the sensitivity is represented by the characteristic noise strain amplitude, $\tilde{h}_n(f)\equiv\sqrt{fS_n(f)}$, which is a unitless quantity derived from the detector noise power spectral density ($|~\tilde{}~|$ denotes frequency-domain quantities). For the Roman EML survey and \textit{Gaia}, the sensitivity is plotted as the minimum instantaneous (i.e., time-domain) GW strain amplitude, $h$, which is estimated based on scaling arguments in Equation~\ref{Eqn:hc}. This choice of representation is motivated by the fact that within the frequency sensitivity range of the Roman EML survey and \textit{Gaia}, we expect to see mostly monochromatic GWs. Finally, we reiterate that this estimate method ignores the telescope duty cycle and does not model the effect of having six separated observing seasons on signals at various frequencies.

In Figure~\ref{fig:hc_cam}, the colored blocks denote example sources within each detector frequency range. In the IPTA frequency range, the yellow block shows the characteristic strain of the expected supermassive black hole binary background \cite{Sesana2008}, $\tilde{h}_c$(SMBHBB). In the frequency range of LISA, we plot the characteristic strain amplitude of an illustrative GW source with $\mchirp=10^6~M_\odot, q=1$ at $D_L=25$ Gpc. At the low-frequency end, this signal is truncated arbitrarily at $3\times 10^{-4}~{\rm{Hz}}$ for visual clarity. At the high-frequency end, it is stopped at the corresponding $f_{\rm ISCO}$. The slope of the signal is $-1/6$, characteristic of the inspiral-stage of a black hole binary GW \cite{Cutler1994,Moore2015}. In the frequency band of the Roman EML survey, we show in red and violet blocks the instantaneous GW strain amplitude with chirp masses $10^{9.7}~M_\odot$ and $10^{7}~M_\odot$ at 50 Mpc, respectively. The starting frequency is set arbitrarily to the low frequency limit of the Roman EML survey, and the signal is cut off at $f_{\rm ISCO}$. The instantaneous amplitude scales as $f^{3/2}$ \cite{Blanchet2014,GWPlotter}. It should be noted that these individual binary source signals are merely illustrative; they do not represent the expected GW source population from binary black hole formation theory.

We observe that were the Roman EML survey able to observe the mean deflection signal, it would outperformed \textit{Gaia} at overlapping frequency ranges. This is mainly due to the high cadence observations. However, the mean-subtraction procedure considerably curbs its expected performance. Nonetheless, its high-cadence observations allow for the detection of $10^{-6}~\rm{Hz}-10^{-5}~\rm{Hz}$ GWs, which are inaccessible by other dedicated GW observatories, such as PTAs and LISA. In this range, possible GW sources include SMBHBs with $\mchirp\sim 10^8~ M_\odot-10^9~M_\odot$ at later stages of the inspiral. Due to the larger GW amplitude, such sources will be more detectable than the same GW population earlier in their inspirals, which are targets of PTAs. At the high frequency range, $10^5~M_\odot$ massive black hole binaries and highly eccentric binaries are at the inspiral stage. LISA, on the other hand, will observe these systems much closer to their coalescence \cite{Banerjee,Banerjee2}. Observing the different stages of this population offers invaluable data for piecing together their evolution process, emphasizing the potential of the astrometry GW detection method.

\subsection{MCMC Sensitivity Threshold Analysis}

GW signals in astrometric measurements can be extracted via Bayesian inference. This analysis framework is demonstrated in \cite{Moore2017}, where the authors implement a signal injection-retrieval study tailored for \textit{Gaia}. Specifically, they consider a set of mock \textit{Gaia} exposures and obtain posterior distributions for seven GW source parameters, plus and cross polarization amplitudes, $h_{+,\times}$, their respective initial phases, $\phi_{+,\times}$, GW frequency, $f$ and two angles describing direction to the GW source, $\vec{q}$ (equivalent to $-\mathbf{p}$ in Section~\ref{AllFishingBoats}). In this paper, we focus on characterizing the intrinsic binary parameters that are detectable from the Roman EML survey data. For this purpose, we fix the extrinsic parameters (i.e., GW phase, polarization angle and source position) and derive limits on the binary chirp mass, $\mchirp$, and luminosity distance, $D_L$, across the Roman EML survey frequency spectrum. Specifically, we set the wave phase, inclination angle, and polarization angle to 0. We also fix the GW source at the zenith position in the Galactic frame, as illustrated in Figure~\ref{fig:dnFoV}. The FoV is modeled as a $0.53\,\deg\times0.53\,\deg$ square centered on the galactic center. Fixing the contribution from phase and positional parameters, either by assigning specific representative values, as we do, or by numerically and analytically marginalizing over them, is also commonly adopted in PTA studies to reduce search space dimensions \citep[see, e.g.,][]{Babak2012,Ellis2012,Moore2015}. 

Under our assumption, we consider the optimal case for detection. As the relative angle between the star position, $\mathbf{n}$, and the GW source position, $\vec{q}$, decreases, signal magnitude decreases accordingly and the detection threshold becomes more stringent. By fixing the GW phase to be 0, we simulate the \textit{a posteriori} analysis, where, after observing at least one deflection cycle, we can determine the deflection amplitude from the entire data set.

Rather than calculating the full posterior distribution from mock data as in Ref.~\cite{Moore2017}, we estimate the detection threshold by computing the likelihood of the signal-present hypothesis for various GW sources assuming we have observed the maximal astrometric deflection from the baseline. The astrometry measurement error is assumed to follow a Gaussian distribution with zero mean and no correlation across time. The standard deviation is the single-exposure, single-source astrometric accuracy. The signal is the mean-subtracted $dn$ in Equation~\ref{eqn:dn}.

For simplicity and computational efficiency, we include only a subset of the expected number of observed stars and consider a single exposure at the maximal deflection. We then scale the results to approximate analysis outcomes with a full mock data set. Specifically, we randomly populate the FoV with 1000 stars. To account for the effect of the expected $10^8$ observed stars, we scale down the astrometric resolution, $\sigma$, by $\sqrt{10^5}$. We make tests using several start counts ranging from $10^3$ to $10^6$, and observe no systematic bias. As Equation~\ref{Eqn:hc} suggests, we simulate the many exposures by scaling down $\sigma$ by $\sqrt{N_m}$. See Appendix~\ref{app:scale} for further justification.

We calculate the likelihood with Markov chain Monte Carlo (MCMC) simulations using the Python package \texttt{emcee} with no injection signal. We determine the $68\%, 95\% ~{\rm{and}}~ 99.7\%$ upper limits on $M_s$. We adopt a flat prior between: $$4.54<\log_{10} M_s~[M_\odot/{\rm{Mpc}}^{3/5}]<11.54 \; ,$$ which is equivalent to flat priors between: $$5.74<\log_{10} \mchirp~\left[M_\odot\right]<9.74$$ and $$-3<\log_{10} D_L~\left[{\rm{Mpc}}\right]<2 \; .$$ 

The upper bound of chirp mass is chosen such that the GW sources are realistic and have significant lifetime within the frequency band of the Roman EML survey. Parameter limits on the luminosity distance and the chirp mass are chosen to produce sufficiently strong signals in light of the theoretically calculated sensitivity curve. 

\begin{figure*}[!htb]
    \includegraphics[width=.7\textwidth]{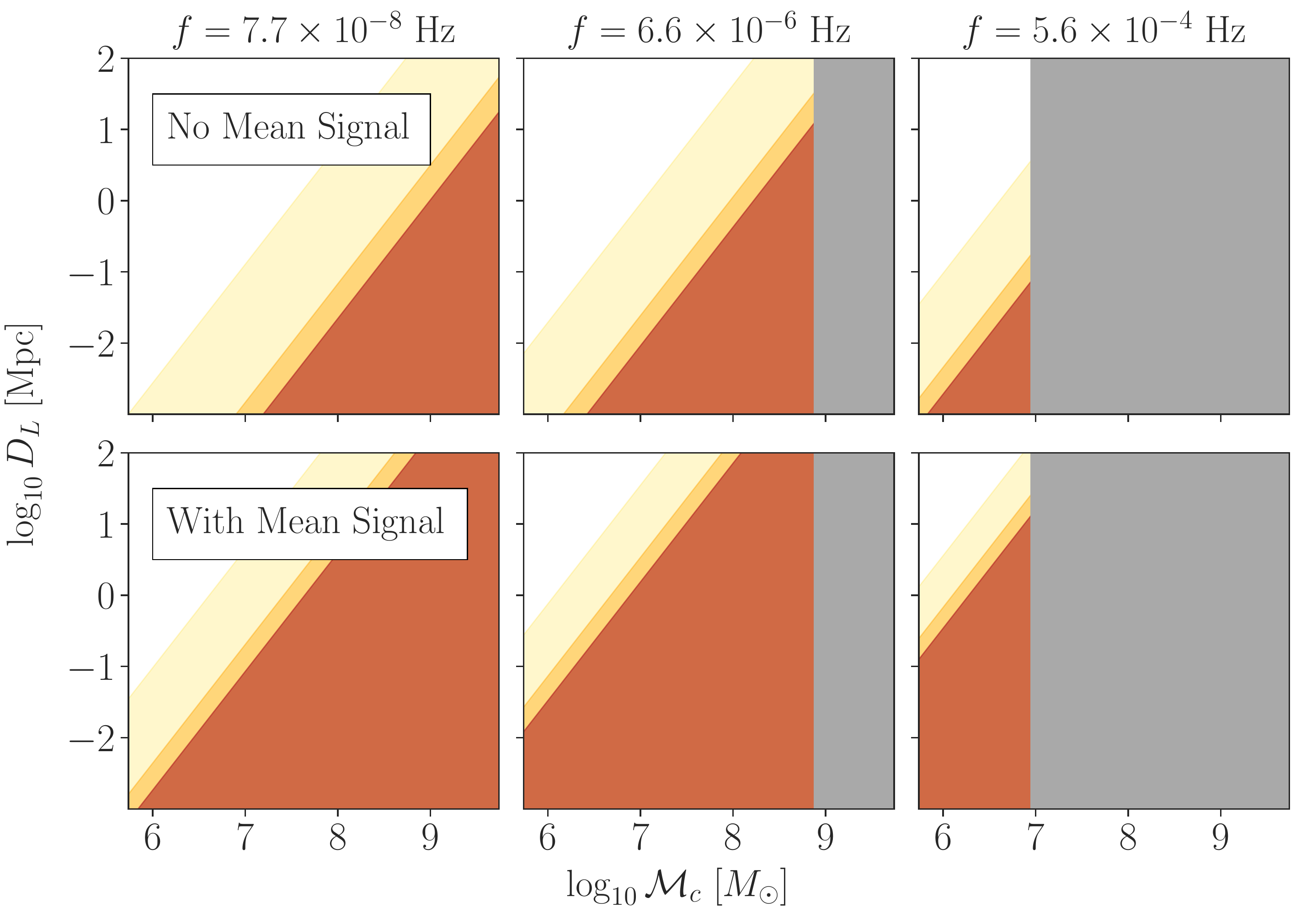}
    \caption{\small{Detection sensitivity of the Roman EML Survey in $\log_{10} \mchirp-\log_{10} D_L$ space at multiple fixed frequency. Colors indicate detection thresholds at different confidence levels. \textit{Top Row:} sensitivity when signals are mean-subtracted. \textit{Bottom Row:} Sensitivity when the full signal is observable. For the three columns, the GW frequencies are fixed to be $7.7\times10^{-8}~{\rm Hz},6.6\times10^{-6}~{\rm Hz}$ and $5.6\times10^{-4}~{\rm Hz}$, respectively. In all panels, GW sources that reach the ISCO at the specified frequency or lower are blocked out in gray.}}
    \label{fig:McR_thresh}
\end{figure*}

The detectable $\log_{10} \mchirp-\log_{10} D_L$ parameter space at selected GW frequencies is show in Figure~\ref{fig:McR_thresh}. The top row shows the detectable binaries when the mean signal is subtracted, and the bottom row shows those when the full signal can be registered. The columns represent the sensitivity at different GW frequencies. Systems that already reach the ISCO are excluded from the accessible parameter space, since they quickly coalesce afterwards, and our analytical waveform expression in Equation~\ref{eqn:h0} for the inspiral stage no longer captures the actual GW waveform. Specifically, \citep[see, e.g.,][]{Sesana2004}
\begin{equation}
    f_{\rm{ISCO}} = \frac{c^3}{6^{3/2}\pi G}\frac{(q^{2}(1+q))^{3/10}}{\mchirp} \; ,
\end{equation} 
and 
\begin{equation}
    \mchirp(f)_{\rm{max}} = \frac{c^3}{6^{3/2}\pi G}\frac{2^{3/10}}{f} \; ,
\end{equation}
where, since $\mchirp(f)_{\rm{max}}$ is an increasing function of $q$, we set $q=1$.

Figure.~\ref{fig:McR_thresh} shows that, at all frequencies, the detectable parameter space is reduced significantly by subtracting the mean signal, and for GW with frequencies larger than $1\times10^{-6}$ Hz, the parameter space is increasingly affected by the ISCO limit.

\begin{figure}
    \centering
    \includegraphics[width=\columnwidth]{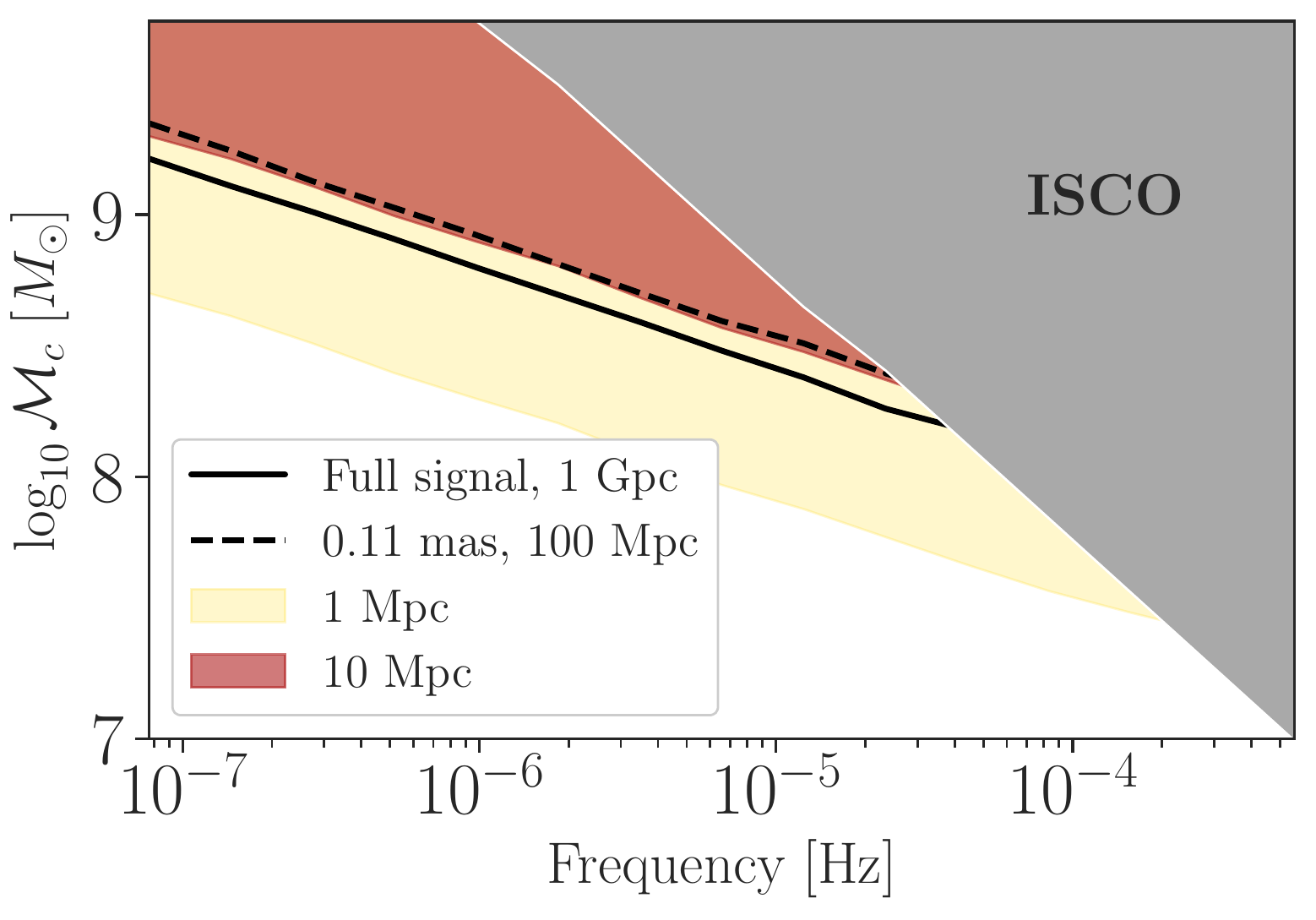}
    \caption{\small{Sensitivity of the Roman EML Survey to $\log_{10} \mchirp$. The upper right corner (shaded gray) excludes massive systems that reach the ISCO at each frequency or lower. Detection sensitivity threshold is represented by the detectable chirp mass at a 2$\sigma$ confidence level at various luminosity distances. Yellow and red blocks show detectable mass ranges at 1 Mpc and 10 Mpc respectively, assuming the mean astrometric deflection is subtracted from the signal. The dashed line shows the detection threshold at 100 Mpc if the astrometric accuracy were to improve to 0.11 mas, equivalent to a factor of 10 improvement in the sensitivity. The solid line shows the sensitivity at 1 Gpc if the mean signal were observable, roughly comparable to a factor of 100 sensitivity improvement. See Section~\ref{OurNestIsBest} for further discussions.}}
    \label{fig:thresh}
\end{figure}

We summarize in Figure~\ref{fig:thresh} the detection threshold across the Roman EML survey frequency band by plotting the $95\%$ upper limit on $\mchirp$ at 1 Mpc and 10 Mpc. Detection thresholds assuming an astrometric accuracy of 0.11 mas or full astrometric deflection signal are also plotted. As expected, the range of the detectable GW sources is limited by the signal strength and intrinsic frequency limits (i.e., $f_{\rm ISCO}$). Between these two competing factors, the ``sweet spot'' frequency with the largest accessible parameter space in the $\log_{10}\mchirp-\log_{10}D_L$ plane is roughly located at $10^{-6}$ Hz. With its current expected performance, the Roman EML survey is sensitive to GWs from massive black hole binaries with $\mchirp> 10^{7.4}~M_\odot$ up to $D_L\sim1$ Mpc; up to $D_L\sim10$ Mpc, binaries with $\mchirp> 10^{8.3}~M_\odot$. Although this threshold excludes many of the interesting GW sources we hope to detect, Figure~\ref{fig:thresh} shows that such sources out to 100 Mpc could be observable if the Roman Space Telescope can achieve a 0.11 mas astrometric accuracy, which is a possible improvement over the currently estimated 1.1 mas.

 \section{Discussion}
 \label{OurNestIsBest}

In this section, we elaborate on the Roman Space Telescope pointing reconstruction strategy and evaluate its impact on GW detection. We then propose recommendations for maximizing the seredipitous GW scientific output from photometric survey instruments. Finally, we review some ongoing and planned surveys and discuss their merits and drawbacks as potential GW probes.

\subsection{Roman Space Telescope Pointing Reconstruction and GS Selection}

Here we expand on the mean subtraction technique discussed in Section \ref{OurFishingBoat} and assess its impact on the reach of the Roman EML survey as a GW probe. 

Prior to launch, 4 to 18 guiding stars (GSs) will be selected in each observed field \cite{Sanderson2019}. Of the 18 detectors of the Roman Space Telescope, each contains at most one guiding star. These stars are likely to be bright, and their precise absolute positions and proper motion will be available in external catalogs, e.g., in the \textit{Gaia} catalog \cite{Sanderson2019}. Their astrometric solution in the Roman Space Telescope operational epoch is extrapolated from the external catalog measurement \citep[a similar procedure to study proper motions of galactic bulge stars is described in][]{Clarkson2008}. The absolute astrometry of all stars in the FoV is then obtained in post-processing by simultaneously fitting the GSs to their extrapolated positions.

As argued in Section~\ref{OurFishingBoat}, this tracking process will likely absorb a mean displacement signal within the FoV. Specifically for the Roman Space Telescope, this will be the mean deflections of the GSs. Though the choice of GSs is not yet available, we can gauge the effect of GS selection by repeating the MCMC study but subtracting only the mean of the GSs. For simplicity, we study two cases with 4 and 16 GSs.
In each case, we model the detectors as square blocks that completely fill the FoV (i.e. no gaps, etc.) and place one GS in each of the square blocks. The position of the GS within each detector is then randomly chosen. We find that different GS choices only lead to $<1\%$ variation in the upper limit confidence value, and having fewer GSs gives larger variations. Thus, our strategy to subtract the mean of all stars serves as a good reference regardless of the mission specifics. 

This mean-subtraction process significantly reduces the effective signal, and the sensitivity level is generally two orders of magnitude lower than the full-signal scenario. Figure~\ref{fig:thresh} shows that the Roman EML survey is most sensitive to very massive binaries $\left(\sim10^8~M_\odot\right)$ at close distances $\left(\sim1~\rm{Mpc}\right)$. Since this is physically unlikely, the Roman EML survey with its current design will be limited as a GW probe. In fact, Figure~\ref{fig:hc_cam} shows that the Roman EML survey would have better sensitivity than \textit{Gaia} if the mean signal were to be detectable, in which case the accessible parameter space would be greatly expanded.

This prediction is different from that for \textit{Gaia} in \cite{Moore2017}, since a full-signal analysis is assumed. In the case of \textit{Gaia}, this treatment is warranted since \textit{Gaia} simultaneously observes through two widely separated FoVs and does not need to perform mean subtraction \cite{Gaia2016}. The sampled GW deflection patterns are consequently distinct and cannot be absorbed by the same pointing calibration process. For essentially the same reason, $Gaia$ can measure absolute parallax rather than relative parallax \cite{Lindegren}.

While this outlined strategy is specific to the Roman EML survey, we note that the loss of the mean astrometric deflection signal is a typical feature of photometric surveys. Even though this loss presents a challenge for resolving individual GWs, the sensitivity might be better for joint signals of several GWs. We expect the combined GWs to produce a deflection pattern richer in features, and thus easier to detect. Such signals would come from SMBHBs at the centers of galaxies in the local universe, and the astrometric measurements can be used to study their population statistics. This is analogous to using PTA measurements to constrain the energy density of the stochastic GW background produced by massive black hole mergers across all redshifts \citep[see, e.g.,][]{Pyne1996,vanHaasteren2011,NANOGrav112018}. It is estimated that $\sim 100$ continuous GW sources in the PTA band exist within 225 Mpc \cite{Mingarelli2017}. We may then speculate that a significant number of SMBHBs within our frequency range exist in the local universe, and their joint signal may be above detection threshold for the astrometry method.

\subsection{Optimizing Photometric Surveys for GW}

In this section, we give specific recommendations to maximize the GW detection potential of photometric surveys within the GW frequency gap of existing detection methods. We use the expected performance of the Roman EML survey as a reference point and quantitatively describe a model survey capable of detecting a fiducial target, a $10^7~M_\odot$ binary at $50~\rm{Mpc}$.  

To estimate the required sensitivity, we proceed directly from Figure~\ref{fig:thresh}; the detection threshold is lowered by the same order of magnitude as the increase in effective signal strength. Therefore, to claim a $2\sigma$ detection on GWs from this new fiducial target, the model survey is 100 times more sensitive than the Roman EML survey. In the following sections we discuss ways to achieve this sensitivity.

\subsubsection{Mean-signal Recovery Fraction}

As suggested by the previous subsection, pointing reconstruction strategies determine whether the mean astrometric deflection could be observable, which translates into approximately two orders of magnitude sensitivity difference. Though the mean-subtracted deflection pattern and the full signal pattern differ in both the deflection magnitude averaged over all stars within the FoV and the pattern shape, we only use the average deflection magnitude as an approximate metric to compare sensitivity. 

For our model FoV configuration, the average deflection magnitude after subtracting the mean is roughly 100 times smaller than the average full signal (see, for example, Figure~\ref{fig:dnFoV}), thus the sensitivity is roughly 100 times worse. We may then define a mean-signal recovery fraction to roughly quantify the observable deflection relative to the full signal. For example, a mean-signal recovery fraction of 50\% implies that the average magnitude of the observable deflections after the astrometry solution is half of the full signal magnitude. Consequently, the sensitivity would be roughly 50 times higher than what we obtained from the MCMC study assuming mean-subtracted signals. Ideally, the photometric survey retains nearly all of the mean signal, relaxing the detection threshold by roughly a factor of 100. Such a model survey, with all other parameters similar to the Roman EML survey, can already detect the fiducial GW source. In general, higher recovery fraction allows detection of intrinsically weaker GW sources, such as farther and lighter systems, or the same system but much earlier in its evolution track.

\subsubsection{Astrometric Accuracy}

As shown by Equation~\ref{Eqn:hc}, the strain amplitude threshold is linearly proportional to the astrometric accuracy. All else equivalent, the model survey improves upon the Roman Space Telescope accuracy by at least a factor of 100, giving a single-exposure single-source astrometric resolution better than $11~\mu\rm{as}$. In this work, we assume an astrometric accuracy of 1.1 mas, which is $1/100$th of the detector pixel size \cite{Sanderson2019}. We expect this accuracy to be routinely performed, but it is possible that 0.11 mas can be achieved \cite{JPL}. In this case, the Roman EML survey will be ten times more sensitive and will already be able to detect binaries with $\mchirp>10^{8.3}M_\odot$ within 100 Mpc (see Figure~\ref{fig:thresh}).

For comparison, the astrometric accuracy of \textit{Gaia} is $0.1\sim2$ mas (for $G=17$ and $G=21$ stars, respectively) \cite{GaiaDR2_2018}. The expected imaging resolution of the Square Kilometer Array (SKA)\footnote{\texttt{https://www.skatelescope.org/the-ska-project/}} \cite{Weltman2020} at 12.5 GHz is 0.04 arcsec \cite{SKAinfo}; assuming a fractional position error requirement smaller than 1\% \cite{SKAreq}, the SKA can achieve an astrometric accuracy better than 0.4 mas.

\subsubsection{Number of Stars}
The statistical advantage of observing more stars ($\propto \sqrt{N_s}$) is stated in Equation~\ref{Eqn:hc}. This number can be expressed as 
\begin{equation}
\begin{split}
    N_s &= \int \alpha\left(\vec{r}\right)\rho\left(L,\vec{r}\right)dL~d^3\vec{r}_{\rm{obs}}\\&\approx A\int \alpha(\theta_0,\phi_0,r)\rho(L,\theta_0,\phi_0,r)r^2dL~dr \; ,
\end{split}
\end{equation}
where $\alpha\left(\vec{r}\right)$ is the effective detectable fraction after photon loss during propagation (e.g., dust absorption, crowding effect, etc.), and $\rho\left(L,\vec{r}\right)$ is the population density of luminosity-$L$ stars. $A$ is the covered angular area. In the second equality, we assume small variation of the integrand in the angular directions. Since GW detection requires frequent visit to the same field, we assume a ``deep survey'' mode where the total surveyed angular area is small and this equality is satisfied. 

Evidently, the observational efficiency of telescopes is greatly increased if they can penetrate to further distances per area (i.e., large $\alpha$). Therefore, a telescope with near-infrared filters outperforms one operating in the visible band, as near-infrared photons suffer less absorption by galactic dust along propagation. The optimal choice for the filter wavelength should, however, be balanced between this low-absorption advantage and the large-diffraction effect for long wavelengths, which degrades the point spread function and thus the astrometric accuracy. 

The $\theta_0,\phi_0$ dependence suggests the importance of pointing directions. Specifically, surveys pointing toward the galactic center have larger $\rho$ for fixed distance and luminosity. For magnitude limited surveys, this implies a larger number of observed stars. Conversely, surveys in high latitude regions are less advantageous since they observe fewer stars above certain magnitude limits. For comparison, the stellar density down to $H(AB) = 20$ mag at Galactic Latitude of $60\deg$ is approximately 3000 stars/deg$^2$ \cite{Euclid2012}.

It is also intuitive that a larger FoV leads to more observed stars, all else equivalent. Therefore, the model survey will have comparable bands and pointing directions to the Roman Space Telescope during its EML survey, but with a 100 times larger ($\sim 200\deg^2$) survey area. 

\subsubsection{FoV Size}
The effect of increasing the FoV size is partially degenerate with increasing the survey area, but it also allows for a larger deflection residual after mean subtraction. Specifically, the subtracted mean decreases as the variation across the FoV at each exposure becomes more significant. The scaling relation between the FoV sidelength, average deflection vector field divergence and the average deflection magnitude is described in Section~\ref{AllFishingBoats}.

The combination of changes both in signal magnitude and pattern will likely be a complex effect that, in general, enhance the sensitivity. In principle, FoV patches can be stitched together to provide a larger effective FoV to include more pattern variation. However, the field-switching process must be exquisitely controlled such that the absorbed mean for each field is approximately the global mean solution in the larger effective FoV. However, due to the very large scale over which the GW-induced deflection pattern varies (on the order of tens of degrees), it is unlikely that future surveys can outperform the Roman EML survey by a factor of 100 through this means alone. 

\subsubsection{Observational Cadence \& Mission Length}

The impact of observational cadence is two-fold: it determines the sensitive frequency range and contributes to the statistical improvement of sensitivity. To complement LISA, therefore, the upper limit frequency should be at least $\sim 10^{-5}~\rm{Hz}$. It follows that an the model survey observes the same patch of sky at least once a day. For sensitivity improvement, the model survey has a longer effective observational time than the six 72-day epochs of the Roman EML survey. For example, a 10-year survey with full duty cycle improves the sensitivity by a factor of 3. 

\subsection{Other Potential Photometric GW Probes}

In this section, we further develop the guidelines for assessing photometric surveys as GW probes. We discuss ground-based and space-based telescopes in turn by pointing out their respective merits and drawbacks as potential GW probes. To the best of our knowledge, all the observatories discussed below would suffer from the limiting mean signal subtraction we discussed above.

A challenge with ground-based telescopes as GW probes lies in their relatively coarse astrometric resolution compared with space telescopes, due to atmospheric perturbation to the signal. For example, the Rubin Observatory\footnote{\texttt{https://www.lsst.org/lsst/}} has a single-exposure astrometric accuracy of $\sim11$ mas \cite{LSST2019}, an order of magnitude larger than that of the Roman Space Telescope. 

This resolution drawback can be partially compensated by a large number of observed stars, large FoV, and great observational flexibility. For instance, the Rubin Observatory is expected to observe a total of $\sim4\times10^9$ stars with a FoV size of $\sim10\deg^2$. Each sky patch is visited $\sim100$ times during its 10-year lifetime. By increasing its observational cadence by a factor of 5 ($\sim 1$ week$^{-1}$ on average), the Rubin Observatory would become sensitive to GWs with $f<1.5\times10^{-6}$ Hz.

This astrometric method can also be applied to high-resolution radio telescopes, such as the SKA and the Next Generation Very Large Array (ngVLA)\footnote{\texttt{https://ngvla.nrao.edu/}} \cite{ngVLA2019}. As discussed, SKA can achieve an astrometric accuracy better than 0.4 mas. The ngVLA features a maximum baseline resolution as small as $0.17~\rm{mas}$ at $41~\rm{GHz}$. It is also estimated that a large number of quasars can be observed in the radio band \citep[see, e.g., $\sim10^6$ in][]{Jaffe2004}, which can serve as GW detectors instead of stars. Taking the SKA as an example, the relatively smaller number of observed quasars compared with stars observed by \textit{Gaia} can potentially be compensated by a more frequent observation schedule to give similar performance at a higher frequency. Increasing the exposure time would also directly increase the number of detected quasars. Specifically, an SKA survey taking measurements every 40 minutes has a GW frequency band similar to that of the Roman EML survey ($f<2\times10^{-4}$ Hz). To the authors' knowledge, there is currently no high-cadence survey planned. 

Free from atmospheric effects, space-based telescopes can potentially observe a great number of stars to excellent precision. As an example, the ESA mission Euclid\footnote{\texttt{https://sci.esa.int/web/euclid/home}} is equipped with a near-infrared photometer with a $\sim1$ mas single-exposure astrometric accuracy \cite{Euclid2019}, similar to the expected performance of the Roman Space Telescope. The HabEx Workhorse Camera (HWC) onboard the Habitable Exoplanet Observatory (HabEx)\footnote{\texttt{https://www.jpl.nasa.gov/habex/}} is expected to have similar, if not better, angular resolution to the Roman Space Telescope, albeit with a much smaller FoV \cite{Habex2020}. 

The deciding factors then become the observed fields and observational cadence. Unlike the Roman Space Telescope, HabEx is not designed as a survey instrument; instead, it focuses on characterizing a handful of targets in great detail, and thus will not be suitable for our purpose. While Euclid does feature a deep survey, with $40\deg^2$ of sky observed every 15 days \cite{Euclid2019}, these fields are close to the ecliptic pole with low stellar density. However, a high-cadence survey in its extended mission lifetime, following the recommendations we outline, could contribute meaningfully to GW detection. 

 \section{Concluding Remarks}
 \label{conclusion}

In this paper, we show how to use a photometric survey as a GW probe that uniquely bridges the GW frequency spectrum gap between existing detection methods. It is also not required that such photometric surveys are designed specifically to provide absolute astrometric solutions. We demonstrate that relative astrometric deflections of observed stars within the FoV already allow for GW detection, albeit at a lower sensitivity level. We discuss key factors that determine sensitivity. We then assess the potential of the Roman EML survey in its current definition as a GW probe. In Section~\ref{OurNestIsBest}, we make recommendations for maximizing the GW scientific output of photometric surveys and quantify the desirable performance via a model survey. Finally, we review existing and planned photometric surveys, and discuss their relative strengths and drawbacks as potential GW probes.

We note that our analysis can be refined in several ways. For instance, we have yet to explicitly include stellar proper motion in our simulation, which can in theory be subtracted via quadratic fitting. Such proper motions may even be correlated across the FoV, if, for example, open clusters are present. However, we expect these motions to have limited impact on the GW sensitivity once we consider the signal variation over time. Especially for high-frequency GWs, their oscillatory nature leaves a distinct signature from physical proper motion over long timescales. 

We could also model the seven fields of the Roman EML survey jointly. A combined analysis of the data from all fields might amount to having a larger effective FoV, should the temporal and pointing accuracy during the field-switching process allow. Incorporating the GW frequency evolution could also enhance sensitivity. 

The recommendations in Section~\ref{OurFishingBoat} should serve as a reference for maximizing GW science from future photometric surveys. The current expected performance of the Roman Space Telescope could make detecting individual GWs a challenge. However, with some luck and a novel pointing reconstruction strategy, we may yet detect individual GWs with the Roman Space Telescope. 

\acknowledgements
We thank the anonymous referees for their helpful comments and suggestions that improve this manuscript. We are deeply grateful for discussions with Jeff Kruk, Scott Gaudi, and Davy Kirkpatric on the Roman Space Telescope, as well as Robert Lupton on the Rubin Observatory and Joseph Lazio and Steven Myers on the ngVLA. We thank Todd Gaier for a discussion that inspired this work. YW would like to thank the David and Ellen Lee Distinguished Fellowship for support during this research. Part of this work was done at Jet Propulsion Laboratory, California Institute of Technology, under a contract with the National Aeronautics and Space Administration. This work was supported by NASA grant 15-WFIRST15-0008 \textit{Cosmology with the High Latitude Survey} Roman Science Investigation Team (SIT).

\textit{Software:} astropy \cite{astropy}, astroquery \cite{astroquery}, emcee \cite{Foreman-Mackey2013}, matplotlib \cite{matplotlib}, numpy \cite{numpy}, scipy \cite{scipy}

\bibliographystyle{apsrev4-2}
\bibliography{ref}

\begin{thebibliography}{62}%
\makeatletter
\providecommand \@ifxundefined [1]{%
 \@ifx{#1\undefined}
}%
\providecommand \@ifnum [1]{%
 \ifnum #1\expandafter \@firstoftwo
 \else \expandafter \@secondoftwo
 \fi
}%
\providecommand \@ifx [1]{%
 \ifx #1\expandafter \@firstoftwo
 \else \expandafter \@secondoftwo
 \fi
}%
\providecommand \natexlab [1]{#1}%
\providecommand \enquote  [1]{``#1''}%
\providecommand \bibnamefont  [1]{#1}%
\providecommand \bibfnamefont [1]{#1}%
\providecommand \citenamefont [1]{#1}%
\providecommand \href@noop [0]{\@secondoftwo}%
\providecommand \href [0]{\begingroup \@sanitize@url \@href}%
\providecommand \@href[1]{\@@startlink{#1}\@@href}%
\providecommand \@@href[1]{\endgroup#1\@@endlink}%
\providecommand \@sanitize@url [0]{\catcode `\\12\catcode `\$12\catcode
  `\&12\catcode `\#12\catcode `\^12\catcode `\_12\catcode `\%12\relax}%
\providecommand \@@startlink[1]{}%
\providecommand \@@endlink[0]{}%
\providecommand \url  [0]{\begingroup\@sanitize@url \@url }%
\providecommand \@url [1]{\endgroup\@href {#1}{\urlprefix }}%
\providecommand \urlprefix  [0]{URL }%
\providecommand \Eprint [0]{\href }%
\providecommand \doibase [0]{https://doi.org/}%
\providecommand \selectlanguage [0]{\@gobble}%
\providecommand \bibinfo  [0]{\@secondoftwo}%
\providecommand \bibfield  [0]{\@secondoftwo}%
\providecommand \translation [1]{[#1]}%
\providecommand \BibitemOpen [0]{}%
\providecommand \bibitemStop [0]{}%
\providecommand \bibitemNoStop [0]{.\EOS\space}%
\providecommand \EOS [0]{\spacefactor3000\relax}%
\providecommand \BibitemShut  [1]{\csname bibitem#1\endcsname}%
\let\auto@bib@innerbib\@empty
\bibitem [{\citenamefont {{The LIGO Scientific Collaboration}}\ and\
  \citenamefont {{the Virgo Collaboration}}(2020)}]{GW190412}%
  \BibitemOpen
  \bibfield  {author} {\bibinfo {author} {\bibnamefont {{The LIGO Scientific
  Collaboration}}}and\ \bibinfo {author} {\bibnamefont {{the Virgo
  Collaboration}}},\ }\href@noop {} {\bibfield  {journal} {\bibinfo  {journal}
  {arXiv e-prints}\ ,\ \bibinfo {eid} {arXiv:2004.08342}} (\bibinfo {year}
  {2020})},\ \Eprint {https://arxiv.org/abs/2004.08342} {arXiv:2004.08342
  [astro-ph.HE]} \BibitemShut {NoStop}%
\bibitem [{\citenamefont {{Abbott}}\ \emph {et~al.}(2017)\citenamefont
  {{Abbott}} \emph {et~al.}}]{GW170817}%
  \BibitemOpen
  \bibfield  {author} {\bibinfo {author} {\bibfnamefont {B.~P.}\ \bibnamefont
  {{Abbott}}} \emph {et~al.} (\bibinfo {collaboration} {{LIGO Scientific
  Collaboration} and {Virgo Collaboration}}),\ }\href
  {https://doi.org/10.1103/PhysRevLett.119.161101} {\bibfield  {journal}
  {\bibinfo  {journal} {\prl}\ }\textbf {\bibinfo {volume} {119}},\ \bibinfo
  {eid} {161101} (\bibinfo {year} {2017})},\ \Eprint
  {https://arxiv.org/abs/1710.05832} {arXiv:1710.05832 [gr-qc]} \BibitemShut
  {NoStop}%
\bibitem [{\citenamefont {{Aasi}}\ \emph {et~al.}(2015)\citenamefont {{Aasi}}
  \emph {et~al.}}]{Aasi2015}%
  \BibitemOpen
  \bibfield  {author} {\bibinfo {author} {\bibfnamefont {J.}~\bibnamefont
  {{Aasi}}} \emph {et~al.} (\bibinfo {collaboration} {LIGO Scientific
  Collaboration}),\ }\href {https://doi.org/10.1088/0264-9381/32/7/074001}
  {\bibfield  {journal} {\bibinfo  {journal} {Classical and Quantum Gravity}\
  }\textbf {\bibinfo {volume} {32}},\ \bibinfo {eid} {074001} (\bibinfo {year}
  {2015})},\ \Eprint {https://arxiv.org/abs/1411.4547} {arXiv:1411.4547
  [gr-qc]} \BibitemShut {NoStop}%
\bibitem [{\citenamefont {{Kuns}}\ \emph {et~al.}(2019)\citenamefont {{Kuns}},
  \citenamefont {{Yu}}, \citenamefont {{Chen}},\ and\ \citenamefont
  {{Adhikari}}}]{TianGO2019}%
  \BibitemOpen
  \bibfield  {author} {\bibinfo {author} {\bibfnamefont {K.~A.}\ \bibnamefont
  {{Kuns}}}, \bibinfo {author} {\bibfnamefont {H.}~\bibnamefont {{Yu}}},
  \bibinfo {author} {\bibfnamefont {Y.}~\bibnamefont {{Chen}}}, and\ \bibinfo
  {author} {\bibfnamefont {R.~X.}\ \bibnamefont {{Adhikari}}},\ }\href@noop {}
  {\bibfield  {journal} {\bibinfo  {journal} {arXiv e-prints}\ ,\ \bibinfo
  {eid} {arXiv:1908.06004}} (\bibinfo {year} {2019})},\ \Eprint
  {https://arxiv.org/abs/1908.06004} {arXiv:1908.06004 [gr-qc]} \BibitemShut
  {NoStop}%
\bibitem [{\citenamefont {{Kawamura}}\ \emph {et~al.}(2019)\citenamefont
  {{Kawamura}}, \citenamefont {{Nakamura}}, \citenamefont {{Ando}},
  \citenamefont {{Seto}}, \citenamefont {{Akutsu}} \emph
  {et~al.}}]{DECIGO2019}%
  \BibitemOpen
  \bibfield  {author} {\bibinfo {author} {\bibfnamefont {S.}~\bibnamefont
  {{Kawamura}}}, \bibinfo {author} {\bibfnamefont {T.}~\bibnamefont
  {{Nakamura}}}, \bibinfo {author} {\bibfnamefont {M.}~\bibnamefont {{Ando}}},
  \bibinfo {author} {\bibfnamefont {N.}~\bibnamefont {{Seto}}}, \bibinfo
  {author} {\bibfnamefont {T.}~\bibnamefont {{Akutsu}}},  \emph {et~al.},\
  }\href {https://doi.org/10.1142/S0218271818450013} {\bibfield  {journal}
  {\bibinfo  {journal} {International Journal of Modern Physics D}\ }\textbf
  {\bibinfo {volume} {28}},\ \bibinfo {eid} {1845001} (\bibinfo {year}
  {2019})}\BibitemShut {NoStop}%
\bibitem [{\citenamefont {{Amaro-Seoane}}\ \emph {et~al.}(2017)\citenamefont
  {{Amaro-Seoane}}, \citenamefont {{Audley}}, \citenamefont {{Babak}},
  \citenamefont {{Baker}}, \citenamefont {{Barausse}} \emph {et~al.}}]{LISAL3}%
  \BibitemOpen
  \bibfield  {author} {\bibinfo {author} {\bibfnamefont {P.}~\bibnamefont
  {{Amaro-Seoane}}}, \bibinfo {author} {\bibfnamefont {H.}~\bibnamefont
  {{Audley}}}, \bibinfo {author} {\bibfnamefont {S.}~\bibnamefont {{Babak}}},
  \bibinfo {author} {\bibfnamefont {J.}~\bibnamefont {{Baker}}}, \bibinfo
  {author} {\bibfnamefont {E.}~\bibnamefont {{Barausse}}},  \emph {et~al.},\
  }\href@noop {} {\bibfield  {journal} {\bibinfo  {journal} {arXiv e-prints}\
  ,\ \bibinfo {eid} {arXiv:1702.00786}} (\bibinfo {year} {2017})},\ \Eprint
  {https://arxiv.org/abs/1702.00786} {arXiv:1702.00786 [astro-ph.IM]}
  \BibitemShut {NoStop}%
\bibitem [{\citenamefont {{Luo}}\ \emph {et~al.}(2016)\citenamefont {{Luo}},
  \citenamefont {{Chen}}, \citenamefont {{Duan}}, \citenamefont {{Gong}},
  \citenamefont {{Hu}}, \citenamefont {{Ji}}, \citenamefont {{Liu}},
  \citenamefont {{Mei}}, \citenamefont {{Milyukov}}, \citenamefont {{Sazhin}},
  \citenamefont {{Shao}}, \citenamefont {{Toth}}, \citenamefont {{Tu}},
  \citenamefont {{Wang}}, \citenamefont {{Wang}}, \citenamefont {{Yeh}},
  \citenamefont {{Zhan}}, \citenamefont {{Zhang}}, \citenamefont {{Zharov}},\
  and\ \citenamefont {{Zhou}}}]{TianQin2016}%
  \BibitemOpen
  \bibfield  {author} {\bibinfo {author} {\bibfnamefont {J.}~\bibnamefont
  {{Luo}}}, \bibinfo {author} {\bibfnamefont {L.-S.}\ \bibnamefont {{Chen}}},
  \bibinfo {author} {\bibfnamefont {H.-Z.}\ \bibnamefont {{Duan}}}, \bibinfo
  {author} {\bibfnamefont {Y.-G.}\ \bibnamefont {{Gong}}}, \bibinfo {author}
  {\bibfnamefont {S.}~\bibnamefont {{Hu}}}, \bibinfo {author} {\bibfnamefont
  {J.}~\bibnamefont {{Ji}}}, \bibinfo {author} {\bibfnamefont {Q.}~\bibnamefont
  {{Liu}}}, \bibinfo {author} {\bibfnamefont {J.}~\bibnamefont {{Mei}}},
  \bibinfo {author} {\bibfnamefont {V.}~\bibnamefont {{Milyukov}}}, \bibinfo
  {author} {\bibfnamefont {M.}~\bibnamefont {{Sazhin}}}, \bibinfo {author}
  {\bibfnamefont {C.-G.}\ \bibnamefont {{Shao}}}, \bibinfo {author}
  {\bibfnamefont {V.~T.}\ \bibnamefont {{Toth}}}, \bibinfo {author}
  {\bibfnamefont {H.-B.}\ \bibnamefont {{Tu}}}, \bibinfo {author}
  {\bibfnamefont {Y.}~\bibnamefont {{Wang}}}, \bibinfo {author} {\bibfnamefont
  {Y.}~\bibnamefont {{Wang}}}, \bibinfo {author} {\bibfnamefont {H.-C.}\
  \bibnamefont {{Yeh}}}, \bibinfo {author} {\bibfnamefont {M.-S.}\ \bibnamefont
  {{Zhan}}}, \bibinfo {author} {\bibfnamefont {Y.}~\bibnamefont {{Zhang}}},
  \bibinfo {author} {\bibfnamefont {V.}~\bibnamefont {{Zharov}}}, and\ \bibinfo
  {author} {\bibfnamefont {Z.-B.}\ \bibnamefont {{Zhou}}},\ }\href
  {https://doi.org/10.1088/0264-9381/33/3/035010} {\bibfield  {journal}
  {\bibinfo  {journal} {Classical and Quantum Gravity}\ }\textbf {\bibinfo
  {volume} {33}},\ \bibinfo {eid} {035010} (\bibinfo {year} {2016})},\ \Eprint
  {https://arxiv.org/abs/1512.02076} {arXiv:1512.02076 [astro-ph.IM]}
  \BibitemShut {NoStop}%
\bibitem [{\citenamefont {{Science Instrument List}}\ \emph
  {et~al.}(2020)\citenamefont {{Science Instrument List}}, \citenamefont {{:}},
  \citenamefont {{Buikema}}, \citenamefont {{Cahillane}}, \citenamefont
  {{Mansell}}, \citenamefont {{Blair}}, \citenamefont {{Abbott}}, \citenamefont
  {{Adams}}, \citenamefont {{Adhikari}}, \citenamefont {{Ananyeva}},
  \citenamefont {{Appert}}, \citenamefont {{Arai}}, \citenamefont {{Areeda}},
  \citenamefont {{Asali}}, \citenamefont {{Aston}}, \citenamefont {{Austin}},
  \citenamefont {{Baer}}, \citenamefont {{Ball}}, \citenamefont {{Ballmer}},
  \citenamefont {{Banagiri}}, \citenamefont {{Barker}}, \citenamefont
  {{Barsotti}}, \citenamefont {{Bartlett}}, \citenamefont {{Berger}},
  \citenamefont {{Betzwieser}}, \citenamefont {{Bhattacharjee}}, \citenamefont
  {{Billingsley}}, \citenamefont {{Biscans}} \emph {et~al.}}]{O3noise2020}%
  \BibitemOpen
  \bibfield  {author} {\bibinfo {author} {\bibfnamefont {L.}~\bibnamefont
  {{Science Instrument List}}}, \bibinfo {author} {\bibnamefont {{:}}},
  \bibinfo {author} {\bibfnamefont {A.}~\bibnamefont {{Buikema}}}, \bibinfo
  {author} {\bibfnamefont {C.}~\bibnamefont {{Cahillane}}}, \bibinfo {author}
  {\bibfnamefont {G.~L.}\ \bibnamefont {{Mansell}}}, \bibinfo {author}
  {\bibfnamefont {C.~D.}\ \bibnamefont {{Blair}}}, \bibinfo {author}
  {\bibfnamefont {R.}~\bibnamefont {{Abbott}}}, \bibinfo {author}
  {\bibfnamefont {C.}~\bibnamefont {{Adams}}}, \bibinfo {author} {\bibfnamefont
  {R.~X.}\ \bibnamefont {{Adhikari}}}, \bibinfo {author} {\bibfnamefont
  {A.}~\bibnamefont {{Ananyeva}}}, \bibinfo {author} {\bibfnamefont
  {S.}~\bibnamefont {{Appert}}}, \bibinfo {author} {\bibfnamefont
  {K.}~\bibnamefont {{Arai}}}, \bibinfo {author} {\bibfnamefont {J.~S.}\
  \bibnamefont {{Areeda}}}, \bibinfo {author} {\bibfnamefont {Y.}~\bibnamefont
  {{Asali}}}, \bibinfo {author} {\bibfnamefont {S.~M.}\ \bibnamefont
  {{Aston}}}, \bibinfo {author} {\bibfnamefont {C.}~\bibnamefont {{Austin}}},
  \bibinfo {author} {\bibfnamefont {A.~M.}\ \bibnamefont {{Baer}}}, \bibinfo
  {author} {\bibfnamefont {M.}~\bibnamefont {{Ball}}}, \bibinfo {author}
  {\bibfnamefont {S.~W.}\ \bibnamefont {{Ballmer}}}, \bibinfo {author}
  {\bibfnamefont {S.}~\bibnamefont {{Banagiri}}}, \bibinfo {author}
  {\bibfnamefont {D.}~\bibnamefont {{Barker}}}, \bibinfo {author}
  {\bibfnamefont {L.}~\bibnamefont {{Barsotti}}}, \bibinfo {author}
  {\bibfnamefont {J.}~\bibnamefont {{Bartlett}}}, \bibinfo {author}
  {\bibfnamefont {B.~K.}\ \bibnamefont {{Berger}}}, \bibinfo {author}
  {\bibfnamefont {J.}~\bibnamefont {{Betzwieser}}}, \bibinfo {author}
  {\bibfnamefont {D.}~\bibnamefont {{Bhattacharjee}}}, \bibinfo {author}
  {\bibfnamefont {G.}~\bibnamefont {{Billingsley}}}, \bibinfo {author}
  {\bibfnamefont {S.}~\bibnamefont {{Biscans}}},  \emph {et~al.},\ }\href@noop
  {} {\bibfield  {journal} {\bibinfo  {journal} {arXiv e-prints}\ ,\ \bibinfo
  {eid} {arXiv:2008.01301}} (\bibinfo {year} {2020})},\ \Eprint
  {https://arxiv.org/abs/2008.01301} {arXiv:2008.01301 [astro-ph.IM]}
  \BibitemShut {NoStop}%
\bibitem [{\citenamefont {{Thorpe}}\ \emph {et~al.}(2019)\citenamefont
  {{Thorpe}}, \citenamefont {{Ziemer}}, \citenamefont {{Thorpe}}, \citenamefont
  {{Livas}}, \citenamefont {{Conklin}}, \citenamefont {{Caldwell}},
  \citenamefont {{Berti}}, \citenamefont {{McWilliams}}, \citenamefont
  {{Stebbins}}, \citenamefont {{Shoemaker}}, \citenamefont {{Ferrara}},
  \citenamefont {{Larson}} \emph {et~al.}}]{Thorpe2019}%
  \BibitemOpen
  \bibfield  {author} {\bibinfo {author} {\bibfnamefont {J.~I.}\ \bibnamefont
  {{Thorpe}}}, \bibinfo {author} {\bibfnamefont {J.}~\bibnamefont {{Ziemer}}},
  \bibinfo {author} {\bibfnamefont {I.}~\bibnamefont {{Thorpe}}}, \bibinfo
  {author} {\bibfnamefont {J.}~\bibnamefont {{Livas}}}, \bibinfo {author}
  {\bibfnamefont {J.~W.}\ \bibnamefont {{Conklin}}}, \bibinfo {author}
  {\bibfnamefont {R.}~\bibnamefont {{Caldwell}}}, \bibinfo {author}
  {\bibfnamefont {E.}~\bibnamefont {{Berti}}}, \bibinfo {author} {\bibfnamefont
  {S.~T.}\ \bibnamefont {{McWilliams}}}, \bibinfo {author} {\bibfnamefont
  {R.}~\bibnamefont {{Stebbins}}}, \bibinfo {author} {\bibfnamefont
  {D.}~\bibnamefont {{Shoemaker}}}, \bibinfo {author} {\bibfnamefont {E.~C.}\
  \bibnamefont {{Ferrara}}}, \bibinfo {author} {\bibfnamefont {S.~L.}\
  \bibnamefont {{Larson}}},  \emph {et~al.},\ }in\ \href@noop {} {\emph
  {\bibinfo {booktitle} {\baas}}},\ Vol.~\bibinfo {volume} {51}\ (\bibinfo
  {year} {2019})\ p.~\bibinfo {pages} {77}\BibitemShut {NoStop}%
\bibitem [{\citenamefont {{Moore}}\ \emph
  {et~al.}(2015{\natexlab{a}})\citenamefont {{Moore}}, \citenamefont
  {{Taylor}},\ and\ \citenamefont {{Gair}}}]{Moore2015}%
  \BibitemOpen
  \bibfield  {author} {\bibinfo {author} {\bibfnamefont {C.~J.}\ \bibnamefont
  {{Moore}}}, \bibinfo {author} {\bibfnamefont {S.~R.}\ \bibnamefont
  {{Taylor}}}, and\ \bibinfo {author} {\bibfnamefont {J.~R.}\ \bibnamefont
  {{Gair}}},\ }\href {https://doi.org/10.1088/0264-9381/32/5/055004} {\bibfield
   {journal} {\bibinfo  {journal} {Classical and Quantum Gravity}\ }\textbf
  {\bibinfo {volume} {32}},\ \bibinfo {eid} {055004} (\bibinfo {year}
  {2015}{\natexlab{a}})},\ \Eprint {https://arxiv.org/abs/1406.5199}
  {arXiv:1406.5199 [astro-ph.IM]} \BibitemShut {NoStop}%
\bibitem [{\citenamefont {{Babak}}\ and\ \citenamefont
  {{Sesana}}(2012)}]{Babak2012}%
  \BibitemOpen
  \bibfield  {author} {\bibinfo {author} {\bibfnamefont {S.}~\bibnamefont
  {{Babak}}}and\ \bibinfo {author} {\bibfnamefont {A.}~\bibnamefont
  {{Sesana}}},\ }\href {https://doi.org/10.1103/PhysRevD.85.044034} {\bibfield
  {journal} {\bibinfo  {journal} {\prd}\ }\textbf {\bibinfo {volume} {85}},\
  \bibinfo {eid} {044034} (\bibinfo {year} {2012})},\ \Eprint
  {https://arxiv.org/abs/1112.1075} {arXiv:1112.1075 [astro-ph.CO]}
  \BibitemShut {NoStop}%
\bibitem [{\citenamefont {{Ellis}}\ \emph {et~al.}(2012)\citenamefont
  {{Ellis}}, \citenamefont {{Siemens}},\ and\ \citenamefont
  {{Creighton}}}]{Ellis2012}%
  \BibitemOpen
  \bibfield  {author} {\bibinfo {author} {\bibfnamefont {J.~A.}\ \bibnamefont
  {{Ellis}}}, \bibinfo {author} {\bibfnamefont {X.}~\bibnamefont {{Siemens}}},
  and\ \bibinfo {author} {\bibfnamefont {J.~D.~E.}\ \bibnamefont
  {{Creighton}}},\ }\href {https://doi.org/10.1088/0004-637X/756/2/175}
  {\bibfield  {journal} {\bibinfo  {journal} {\apj}\ }\textbf {\bibinfo
  {volume} {756}},\ \bibinfo {eid} {175} (\bibinfo {year} {2012})},\ \Eprint
  {https://arxiv.org/abs/1204.4218} {arXiv:1204.4218 [astro-ph.IM]}
  \BibitemShut {NoStop}%
\bibitem [{\citenamefont {{Taylor}}\ \emph {et~al.}(2016)\citenamefont
  {{Taylor}}, \citenamefont {{Vallisneri}}, \citenamefont {{Ellis}},
  \citenamefont {{Mingarelli}}, \citenamefont {{Lazio}},\ and\ \citenamefont
  {{van Haasteren}}}]{Taylor2016}%
  \BibitemOpen
  \bibfield  {author} {\bibinfo {author} {\bibfnamefont {S.~R.}\ \bibnamefont
  {{Taylor}}}, \bibinfo {author} {\bibfnamefont {M.}~\bibnamefont
  {{Vallisneri}}}, \bibinfo {author} {\bibfnamefont {J.~A.}\ \bibnamefont
  {{Ellis}}}, \bibinfo {author} {\bibfnamefont {C.~M.~F.}\ \bibnamefont
  {{Mingarelli}}}, \bibinfo {author} {\bibfnamefont {T.~J.~W.}\ \bibnamefont
  {{Lazio}}}, and\ \bibinfo {author} {\bibfnamefont {R.}~\bibnamefont {{van
  Haasteren}}},\ }\href {https://doi.org/10.3847/2041-8205/819/1/L6} {\bibfield
   {journal} {\bibinfo  {journal} {\apjl}\ }\textbf {\bibinfo {volume} {819}},\
  \bibinfo {eid} {L6} (\bibinfo {year} {2016})},\ \Eprint
  {https://arxiv.org/abs/1511.05564} {arXiv:1511.05564 [astro-ph.IM]}
  \BibitemShut {NoStop}%
\bibitem [{\citenamefont {{Verbiest}}\ \emph {et~al.}(2016)\citenamefont
  {{Verbiest}}, \citenamefont {{Lentati}}, \citenamefont {{Hobbs}},
  \citenamefont {{van Haasteren}}, \citenamefont {{Demorest}} \emph
  {et~al.}}]{Verbiest2016}%
  \BibitemOpen
  \bibfield  {author} {\bibinfo {author} {\bibfnamefont {J.~P.~W.}\
  \bibnamefont {{Verbiest}}}, \bibinfo {author} {\bibfnamefont
  {L.}~\bibnamefont {{Lentati}}}, \bibinfo {author} {\bibfnamefont
  {G.}~\bibnamefont {{Hobbs}}}, \bibinfo {author} {\bibfnamefont
  {R.}~\bibnamefont {{van Haasteren}}}, \bibinfo {author} {\bibfnamefont
  {P.~B.}\ \bibnamefont {{Demorest}}},  \emph {et~al.},\ }\href
  {https://doi.org/10.1093/mnras/stw347} {\bibfield  {journal} {\bibinfo
  {journal} {\mnras}\ }\textbf {\bibinfo {volume} {458}},\ \bibinfo {pages}
  {1267} (\bibinfo {year} {2016})},\ \Eprint {https://arxiv.org/abs/1602.03640}
  {arXiv:1602.03640 [astro-ph.IM]} \BibitemShut {NoStop}%
\bibitem [{\citenamefont {{Sesana}}\ \emph {et~al.}(2008)\citenamefont
  {{Sesana}}, \citenamefont {{Vecchio}},\ and\ \citenamefont
  {{Colacino}}}]{Sesana2008}%
  \BibitemOpen
  \bibfield  {author} {\bibinfo {author} {\bibfnamefont {A.}~\bibnamefont
  {{Sesana}}}, \bibinfo {author} {\bibfnamefont {A.}~\bibnamefont {{Vecchio}}},
  and\ \bibinfo {author} {\bibfnamefont {C.~N.}\ \bibnamefont {{Colacino}}},\
  }\href {https://doi.org/10.1111/j.1365-2966.2008.13682.x} {\bibfield
  {journal} {\bibinfo  {journal} {\mnras}\ }\textbf {\bibinfo {volume} {390}},\
  \bibinfo {pages} {192} (\bibinfo {year} {2008})},\ \Eprint
  {https://arxiv.org/abs/0804.4476} {arXiv:0804.4476 [astro-ph]} \BibitemShut
  {NoStop}%
\bibitem [{\citenamefont {{Arzoumanian}}\ \emph {et~al.}(2020)\citenamefont
  {{Arzoumanian}}, \citenamefont {{Baker}}, \citenamefont {{Blumer}},
  \citenamefont {{Becsy}}, \citenamefont {{Brazier}}, \citenamefont {{Brook}},
  \citenamefont {{Burke-Spolaor}}, \citenamefont {{Chatterjee}}, \citenamefont
  {{Chen}}, \citenamefont {{Cordes}}, \citenamefont {{Cornish}} \emph
  {et~al.}}]{nanogravGWB}%
  \BibitemOpen
  \bibfield  {author} {\bibinfo {author} {\bibfnamefont {Z.}~\bibnamefont
  {{Arzoumanian}}}, \bibinfo {author} {\bibfnamefont {P.~T.}\ \bibnamefont
  {{Baker}}}, \bibinfo {author} {\bibfnamefont {H.}~\bibnamefont {{Blumer}}},
  \bibinfo {author} {\bibfnamefont {B.}~\bibnamefont {{Becsy}}}, \bibinfo
  {author} {\bibfnamefont {A.}~\bibnamefont {{Brazier}}}, \bibinfo {author}
  {\bibfnamefont {P.~R.}\ \bibnamefont {{Brook}}}, \bibinfo {author}
  {\bibfnamefont {S.}~\bibnamefont {{Burke-Spolaor}}}, \bibinfo {author}
  {\bibfnamefont {S.}~\bibnamefont {{Chatterjee}}}, \bibinfo {author}
  {\bibfnamefont {S.}~\bibnamefont {{Chen}}}, \bibinfo {author} {\bibfnamefont
  {J.~M.}\ \bibnamefont {{Cordes}}}, \bibinfo {author} {\bibfnamefont {N.~J.}\
  \bibnamefont {{Cornish}}},  \emph {et~al.},\ }\href@noop {} {\bibfield
  {journal} {\bibinfo  {journal} {arXiv e-prints}\ ,\ \bibinfo {eid}
  {arXiv:2009.04496}} (\bibinfo {year} {2020})},\ \Eprint
  {https://arxiv.org/abs/2009.04496} {arXiv:2009.04496 [astro-ph.HE]}
  \BibitemShut {NoStop}%
\bibitem [{\citenamefont {{Book}}\ and\ \citenamefont
  {{Flanagan}}(2011)}]{Book2011}%
  \BibitemOpen
  \bibfield  {author} {\bibinfo {author} {\bibfnamefont {L.~G.}\ \bibnamefont
  {{Book}}}and\ \bibinfo {author} {\bibfnamefont {{\'E}.~{\'E}.}\ \bibnamefont
  {{Flanagan}}},\ }\href {https://doi.org/10.1103/PhysRevD.83.024024}
  {\bibfield  {journal} {\bibinfo  {journal} {\prd}\ }\textbf {\bibinfo
  {volume} {83}},\ \bibinfo {eid} {024024} (\bibinfo {year} {2011})},\ \Eprint
  {https://arxiv.org/abs/1009.4192} {arXiv:1009.4192 [astro-ph.CO]}
  \BibitemShut {NoStop}%
\bibitem [{\citenamefont {{Pyne}}\ \emph {et~al.}(1996)\citenamefont {{Pyne}},
  \citenamefont {{Gwinn}}, \citenamefont {{Birkinshaw}}, \citenamefont
  {{Eubanks}},\ and\ \citenamefont {{Matsakis}}}]{Pyne1996}%
  \BibitemOpen
  \bibfield  {author} {\bibinfo {author} {\bibfnamefont {T.}~\bibnamefont
  {{Pyne}}}, \bibinfo {author} {\bibfnamefont {C.~R.}\ \bibnamefont {{Gwinn}}},
  \bibinfo {author} {\bibfnamefont {M.}~\bibnamefont {{Birkinshaw}}}, \bibinfo
  {author} {\bibfnamefont {T.~M.}\ \bibnamefont {{Eubanks}}}, and\ \bibinfo
  {author} {\bibfnamefont {D.~N.}\ \bibnamefont {{Matsakis}}},\ }\href
  {https://doi.org/10.1086/177443} {\bibfield  {journal} {\bibinfo  {journal}
  {\apj}\ }\textbf {\bibinfo {volume} {465}},\ \bibinfo {pages} {566} (\bibinfo
  {year} {1996})},\ \Eprint {https://arxiv.org/abs/astro-ph/9507030}
  {arXiv:astro-ph/9507030 [astro-ph]} \BibitemShut {NoStop}%
\bibitem [{\citenamefont {{Banerjee}}(2020{\natexlab{a}})}]{Banerjee}%
  \BibitemOpen
  \bibfield  {author} {\bibinfo {author} {\bibfnamefont {S.}~\bibnamefont
  {{Banerjee}}},\ }\href {https://doi.org/10.1103/PhysRevD.102.103002}
  {\bibfield  {journal} {\bibinfo  {journal} {\prd}\ }\textbf {\bibinfo
  {volume} {102}},\ \bibinfo {eid} {103002} (\bibinfo {year}
  {2020}{\natexlab{a}})},\ \Eprint {https://arxiv.org/abs/2006.14587}
  {arXiv:2006.14587 [astro-ph.HE]} \BibitemShut {NoStop}%
\bibitem [{\citenamefont {{Banerjee}}(2020{\natexlab{b}})}]{Banerjee2}%
  \BibitemOpen
  \bibfield  {author} {\bibinfo {author} {\bibfnamefont {S.}~\bibnamefont
  {{Banerjee}}},\ }\bibfield  {journal} {\bibinfo  {journal} {\mnras}\ }\href
  {https://doi.org/10.1093/mnras/staa2392} {10.1093/mnras/staa2392} (\bibinfo
  {year} {2020}{\natexlab{b}}),\ \Eprint {https://arxiv.org/abs/2004.07382}
  {arXiv:2004.07382 [astro-ph.HE]} \BibitemShut {NoStop}%
\bibitem [{\citenamefont {{Sesana}}\ and\ \citenamefont
  {{Vecchio}}(2010)}]{sesana_mono1}%
  \BibitemOpen
  \bibfield  {author} {\bibinfo {author} {\bibfnamefont {A.}~\bibnamefont
  {{Sesana}}}and\ \bibinfo {author} {\bibfnamefont {A.}~\bibnamefont
  {{Vecchio}}},\ }\href {https://doi.org/10.1103/PhysRevD.81.104008} {\bibfield
   {journal} {\bibinfo  {journal} {\prd}\ }\textbf {\bibinfo {volume} {81}},\
  \bibinfo {eid} {104008} (\bibinfo {year} {2010})},\ \Eprint
  {https://arxiv.org/abs/1003.0677} {arXiv:1003.0677 [astro-ph.CO]}
  \BibitemShut {NoStop}%
\bibitem [{\citenamefont {{Sesana}}\ \emph {et~al.}(2009)\citenamefont
  {{Sesana}}, \citenamefont {{Vecchio}},\ and\ \citenamefont
  {{Volonteri}}}]{sesana_mono2}%
  \BibitemOpen
  \bibfield  {author} {\bibinfo {author} {\bibfnamefont {A.}~\bibnamefont
  {{Sesana}}}, \bibinfo {author} {\bibfnamefont {A.}~\bibnamefont {{Vecchio}}},
  and\ \bibinfo {author} {\bibfnamefont {M.}~\bibnamefont {{Volonteri}}},\
  }\href {https://doi.org/10.1111/j.1365-2966.2009.14499.x} {\bibfield
  {journal} {\bibinfo  {journal} {\mnras}\ }\textbf {\bibinfo {volume} {394}},\
  \bibinfo {pages} {2255} (\bibinfo {year} {2009})},\ \Eprint
  {https://arxiv.org/abs/0809.3412} {arXiv:0809.3412 [astro-ph]} \BibitemShut
  {NoStop}%
\bibitem [{\citenamefont {{Klioner}}(2018)}]{Klioner2018}%
  \BibitemOpen
  \bibfield  {author} {\bibinfo {author} {\bibfnamefont {S.~A.}\ \bibnamefont
  {{Klioner}}},\ }\href {https://doi.org/10.1088/1361-6382/aa9f57} {\bibfield
  {journal} {\bibinfo  {journal} {Classical and Quantum Gravity}\ }\textbf
  {\bibinfo {volume} {35}},\ \bibinfo {eid} {045005} (\bibinfo {year}
  {2018})},\ \Eprint {https://arxiv.org/abs/1710.11474} {arXiv:1710.11474
  [astro-ph.HE]} \BibitemShut {NoStop}%
\bibitem [{\citenamefont {{Moore}}\ \emph {et~al.}(2017)\citenamefont
  {{Moore}}, \citenamefont {{Mihaylov}}, \citenamefont {{Lasenby}},\ and\
  \citenamefont {{Gilmore}}}]{Moore2017}%
  \BibitemOpen
  \bibfield  {author} {\bibinfo {author} {\bibfnamefont {C.~J.}\ \bibnamefont
  {{Moore}}}, \bibinfo {author} {\bibfnamefont {D.~P.}\ \bibnamefont
  {{Mihaylov}}}, \bibinfo {author} {\bibfnamefont {A.}~\bibnamefont
  {{Lasenby}}}, and\ \bibinfo {author} {\bibfnamefont {G.}~\bibnamefont
  {{Gilmore}}},\ }\href {https://doi.org/10.1103/PhysRevLett.119.261102}
  {\bibfield  {journal} {\bibinfo  {journal} {\prl}\ }\textbf {\bibinfo
  {volume} {119}},\ \bibinfo {eid} {261102} (\bibinfo {year} {2017})},\ \Eprint
  {https://arxiv.org/abs/1707.06239} {arXiv:1707.06239 [astro-ph.IM]}
  \BibitemShut {NoStop}%
\bibitem [{\citenamefont {{Agrawal}}\ \emph {et~al.}(2019)\citenamefont
  {{Agrawal}}, \citenamefont {{Okumura}},\ and\ \citenamefont
  {{Futamase}}}]{Agrawal}%
  \BibitemOpen
  \bibfield  {author} {\bibinfo {author} {\bibfnamefont {A.}~\bibnamefont
  {{Agrawal}}}, \bibinfo {author} {\bibfnamefont {T.}~\bibnamefont
  {{Okumura}}}, and\ \bibinfo {author} {\bibfnamefont {T.}~\bibnamefont
  {{Futamase}}},\ }\href {https://doi.org/10.1103/PhysRevD.100.063534}
  {\bibfield  {journal} {\bibinfo  {journal} {\prd}\ }\textbf {\bibinfo
  {volume} {100}},\ \bibinfo {eid} {063534} (\bibinfo {year} {2019})},\ \Eprint
  {https://arxiv.org/abs/1907.02328} {arXiv:1907.02328 [astro-ph.CO]}
  \BibitemShut {NoStop}%
\bibitem [{Rom(2020)}]{Roman}%
  \BibitemOpen
  \href {https://www.stsci.edu/roman/about/science-themes} {\bibinfo {title}
  {{Roman's Key Science Components}}} (\bibinfo {year} {2020})\BibitemShut
  {NoStop}%
\bibitem [{\citenamefont {{Gaudi}}\ \emph {et~al.}(2019)\citenamefont
  {{Gaudi}}, \citenamefont {{Akeson}}, \citenamefont {{Anderson}},
  \citenamefont {{Bachelet}}, \citenamefont {{Bennett}}, \citenamefont
  {{Bhattacharya}}, \citenamefont {{Bozza}}, \citenamefont {{Calchi Novati}},
  \citenamefont {{Henderson}}, \citenamefont {{Johnson}}, \citenamefont
  {{Kruk}}, \citenamefont {{Lu}}, \citenamefont {{Mao}}, \citenamefont
  {{Montet}}, \citenamefont {{Nataf}}, \citenamefont {{Penny}}, \citenamefont
  {{Poleski}}, \citenamefont {{Ranc}}, \citenamefont {{Sahu}}, \citenamefont
  {{Shvartzvald}}, \citenamefont {{Spergel}}, \citenamefont {{Suzuki}},
  \citenamefont {{Stassun}},\ and\ \citenamefont {{Street}}}]{Gaudi2019}%
  \BibitemOpen
  \bibfield  {author} {\bibinfo {author} {\bibfnamefont {B.~S.}\ \bibnamefont
  {{Gaudi}}}, \bibinfo {author} {\bibfnamefont {R.}~\bibnamefont {{Akeson}}},
  \bibinfo {author} {\bibfnamefont {J.}~\bibnamefont {{Anderson}}}, \bibinfo
  {author} {\bibfnamefont {E.}~\bibnamefont {{Bachelet}}}, \bibinfo {author}
  {\bibfnamefont {D.~P.}\ \bibnamefont {{Bennett}}}, \bibinfo {author}
  {\bibfnamefont {A.}~\bibnamefont {{Bhattacharya}}}, \bibinfo {author}
  {\bibfnamefont {V.}~\bibnamefont {{Bozza}}}, \bibinfo {author} {\bibfnamefont
  {S.}~\bibnamefont {{Calchi Novati}}}, \bibinfo {author} {\bibfnamefont
  {C.~B.}\ \bibnamefont {{Henderson}}}, \bibinfo {author} {\bibfnamefont
  {S.~A.}\ \bibnamefont {{Johnson}}}, \bibinfo {author} {\bibfnamefont
  {J.}~\bibnamefont {{Kruk}}}, \bibinfo {author} {\bibfnamefont {J.~R.}\
  \bibnamefont {{Lu}}}, \bibinfo {author} {\bibfnamefont {S.}~\bibnamefont
  {{Mao}}}, \bibinfo {author} {\bibfnamefont {B.~T.}\ \bibnamefont {{Montet}}},
  \bibinfo {author} {\bibfnamefont {D.~M.}\ \bibnamefont {{Nataf}}}, \bibinfo
  {author} {\bibfnamefont {M.~T.}\ \bibnamefont {{Penny}}}, \bibinfo {author}
  {\bibfnamefont {R.}~\bibnamefont {{Poleski}}}, \bibinfo {author}
  {\bibfnamefont {C.}~\bibnamefont {{Ranc}}}, \bibinfo {author} {\bibfnamefont
  {K.}~\bibnamefont {{Sahu}}}, \bibinfo {author} {\bibfnamefont
  {Y.}~\bibnamefont {{Shvartzvald}}}, \bibinfo {author} {\bibfnamefont {D.~N.}\
  \bibnamefont {{Spergel}}}, \bibinfo {author} {\bibfnamefont {D.}~\bibnamefont
  {{Suzuki}}}, \bibinfo {author} {\bibfnamefont {K.~G.}\ \bibnamefont
  {{Stassun}}}, and\ \bibinfo {author} {\bibfnamefont {R.~A.}\ \bibnamefont
  {{Street}}},\ }\href@noop {} {\bibfield  {journal} {\bibinfo  {journal}
  {\baas}\ }\textbf {\bibinfo {volume} {51}},\ \bibinfo {eid} {211} (\bibinfo
  {year} {2019})},\ \Eprint {https://arxiv.org/abs/1903.08986}
  {arXiv:1903.08986 [astro-ph.SR]} \BibitemShut {NoStop}%
\bibitem [{\citenamefont {{WFIRST Astrometry Working Group}}\ \emph
  {et~al.}(2019)\citenamefont {{WFIRST Astrometry Working Group}},
  \citenamefont {{Sanderson}}, \citenamefont {{Bellini}}, \citenamefont
  {{Casertano}}, \citenamefont {{Lu}}, \citenamefont {{Melchior}},
  \citenamefont {{Libralato}}, \citenamefont {{Bennett}}, \citenamefont
  {{Shao}}, \citenamefont {{Rhodes}}, \citenamefont {{Sohn}}, \citenamefont
  {{Malhotra}}, \citenamefont {{Gaudi}}, \citenamefont {{Fall}}, \citenamefont
  {{Nelan}}, \citenamefont {{Guhathakurta}}, \citenamefont {{Anderson}},\ and\
  \citenamefont {{Ho}}}]{Sanderson2019}%
  \BibitemOpen
  \bibfield  {author} {\bibinfo {author} {\bibnamefont {{WFIRST Astrometry
  Working Group}}}, \bibinfo {author} {\bibfnamefont {R.~E.}\ \bibnamefont
  {{Sanderson}}}, \bibinfo {author} {\bibfnamefont {A.}~\bibnamefont
  {{Bellini}}}, \bibinfo {author} {\bibfnamefont {S.}~\bibnamefont
  {{Casertano}}}, \bibinfo {author} {\bibfnamefont {J.~R.}\ \bibnamefont
  {{Lu}}}, \bibinfo {author} {\bibfnamefont {P.}~\bibnamefont {{Melchior}}},
  \bibinfo {author} {\bibfnamefont {M.}~\bibnamefont {{Libralato}}}, \bibinfo
  {author} {\bibfnamefont {D.}~\bibnamefont {{Bennett}}}, \bibinfo {author}
  {\bibfnamefont {M.}~\bibnamefont {{Shao}}}, \bibinfo {author} {\bibfnamefont
  {J.}~\bibnamefont {{Rhodes}}}, \bibinfo {author} {\bibfnamefont {S.~T.}\
  \bibnamefont {{Sohn}}}, \bibinfo {author} {\bibfnamefont {S.}~\bibnamefont
  {{Malhotra}}}, \bibinfo {author} {\bibfnamefont {S.}~\bibnamefont {{Gaudi}}},
  \bibinfo {author} {\bibfnamefont {S.~M.}\ \bibnamefont {{Fall}}}, \bibinfo
  {author} {\bibfnamefont {E.}~\bibnamefont {{Nelan}}}, \bibinfo {author}
  {\bibfnamefont {P.}~\bibnamefont {{Guhathakurta}}}, \bibinfo {author}
  {\bibfnamefont {J.}~\bibnamefont {{Anderson}}}, and\ \bibinfo {author}
  {\bibfnamefont {S.}~\bibnamefont {{Ho}}},\ }\href
  {https://doi.org/10.1117/1.JATIS.5.4.044005} {\bibfield  {journal} {\bibinfo
  {journal} {Journal of Astronomical Telescopes, Instruments, and Systems}\
  }\textbf {\bibinfo {volume} {5}},\ \bibinfo {eid} {044005} (\bibinfo {year}
  {2019})}\BibitemShut {NoStop}%
\bibitem [{\citenamefont {{Anholm}}\ \emph {et~al.}(2009)\citenamefont
  {{Anholm}}, \citenamefont {{Ballmer}}, \citenamefont {{Creighton}},
  \citenamefont {{Price}},\ and\ \citenamefont {{Siemens}}}]{Anholm2009}%
  \BibitemOpen
  \bibfield  {author} {\bibinfo {author} {\bibfnamefont {M.}~\bibnamefont
  {{Anholm}}}, \bibinfo {author} {\bibfnamefont {S.}~\bibnamefont {{Ballmer}}},
  \bibinfo {author} {\bibfnamefont {J.~D.~E.}\ \bibnamefont {{Creighton}}},
  \bibinfo {author} {\bibfnamefont {L.~R.}\ \bibnamefont {{Price}}}, and\
  \bibinfo {author} {\bibfnamefont {X.}~\bibnamefont {{Siemens}}},\ }\href
  {https://doi.org/10.1103/PhysRevD.79.084030} {\bibfield  {journal} {\bibinfo
  {journal} {\prd}\ }\textbf {\bibinfo {volume} {79}},\ \bibinfo {eid} {084030}
  (\bibinfo {year} {2009})},\ \Eprint {https://arxiv.org/abs/0809.0701}
  {arXiv:0809.0701 [gr-qc]} \BibitemShut {NoStop}%
\bibitem [{\citenamefont {{Blanchet}}(2014)}]{Blanchet2014}%
  \BibitemOpen
  \bibfield  {author} {\bibinfo {author} {\bibfnamefont {L.}~\bibnamefont
  {{Blanchet}}},\ }\href {https://doi.org/10.12942/lrr-2014-2} {\bibfield
  {journal} {\bibinfo  {journal} {Living Reviews in Relativity}\ }\textbf
  {\bibinfo {volume} {17}},\ \bibinfo {eid} {2} (\bibinfo {year} {2014})},\
  \Eprint {https://arxiv.org/abs/1310.1528} {arXiv:1310.1528 [gr-qc]}
  \BibitemShut {NoStop}%
\bibitem [{\citenamefont {{Prusti}}\ \emph {et~al.}(2016)\citenamefont
  {{Prusti}} \emph {et~al.}}]{Gaia2016}%
  \BibitemOpen
  \bibfield  {author} {\bibinfo {author} {\bibfnamefont {T.}~\bibnamefont
  {{Prusti}}} \emph {et~al.} (\bibinfo {collaboration} {Gaia Collaboration}),\
  }\href {https://doi.org/10.1051/0004-6361/201629272} {\bibfield  {journal}
  {\bibinfo  {journal} {\aap}\ }\textbf {\bibinfo {volume} {595}},\ \bibinfo
  {eid} {A1} (\bibinfo {year} {2016})},\ \Eprint
  {https://arxiv.org/abs/1609.04153} {arXiv:1609.04153 [astro-ph.IM]}
  \BibitemShut {NoStop}%
\bibitem [{\citenamefont {{Brown}}\ \emph {et~al.}(2018)\citenamefont {{Brown}}
  \emph {et~al.}}]{GaiaDR2_2018}%
  \BibitemOpen
  \bibfield  {author} {\bibinfo {author} {\bibfnamefont {A.~G.~A.}\
  \bibnamefont {{Brown}}} \emph {et~al.} (\bibinfo {collaboration} {Gaia
  Collaboration}),\ }\href {https://doi.org/10.1051/0004-6361/201833051}
  {\bibfield  {journal} {\bibinfo  {journal} {\aap}\ }\textbf {\bibinfo
  {volume} {616}},\ \bibinfo {eid} {A1} (\bibinfo {year} {2018})},\ \Eprint
  {https://arxiv.org/abs/1804.09365} {arXiv:1804.09365 [astro-ph.GA]}
  \BibitemShut {NoStop}%
\bibitem [{\citenamefont {{Baggett}}\ \emph {et~al.}(2004)\citenamefont
  {{Baggett}}, \citenamefont {{MacKenty}}, \citenamefont {{Robberto}},
  \citenamefont {{Kimble}}, \citenamefont {{Hill}}, \citenamefont {{Bushouse}},
  \citenamefont {{Baggett}}, \citenamefont {{Brown}}, \citenamefont {{Hartig}},
  \citenamefont {{Hilbert}}, \citenamefont {{Reid}}, \citenamefont {{Delo}},
  \citenamefont {{Foltz}}, \citenamefont {{Johnson}}, \citenamefont {{Lupie}},
  \citenamefont {{Malumuth}}, \citenamefont {{Pham}}, \citenamefont
  {{Russell}}, \citenamefont {{Waczynski}}, \citenamefont {{Wen}},\ and\
  \citenamefont {{WFC3 Team}}}]{Hubble}%
  \BibitemOpen
  \bibfield  {author} {\bibinfo {author} {\bibfnamefont {S.~M.}\ \bibnamefont
  {{Baggett}}}, \bibinfo {author} {\bibfnamefont {J.~W.}\ \bibnamefont
  {{MacKenty}}}, \bibinfo {author} {\bibfnamefont {M.}~\bibnamefont
  {{Robberto}}}, \bibinfo {author} {\bibfnamefont {R.~A.}\ \bibnamefont
  {{Kimble}}}, \bibinfo {author} {\bibfnamefont {R.~J.}\ \bibnamefont
  {{Hill}}}, \bibinfo {author} {\bibfnamefont {H.}~\bibnamefont {{Bushouse}}},
  \bibinfo {author} {\bibfnamefont {W.}~\bibnamefont {{Baggett}}}, \bibinfo
  {author} {\bibfnamefont {T.}~\bibnamefont {{Brown}}}, \bibinfo {author}
  {\bibfnamefont {G.}~\bibnamefont {{Hartig}}}, \bibinfo {author}
  {\bibfnamefont {B.}~\bibnamefont {{Hilbert}}}, \bibinfo {author}
  {\bibfnamefont {I.~N.}\ \bibnamefont {{Reid}}}, \bibinfo {author}
  {\bibfnamefont {G.}~\bibnamefont {{Delo}}}, \bibinfo {author} {\bibfnamefont
  {R.}~\bibnamefont {{Foltz}}}, \bibinfo {author} {\bibfnamefont {S.~D.}\
  \bibnamefont {{Johnson}}}, \bibinfo {author} {\bibfnamefont {O.}~\bibnamefont
  {{Lupie}}}, \bibinfo {author} {\bibfnamefont {E.}~\bibnamefont {{Malumuth}}},
  \bibinfo {author} {\bibfnamefont {T.}~\bibnamefont {{Pham}}}, \bibinfo
  {author} {\bibfnamefont {A.~M.}\ \bibnamefont {{Russell}}}, \bibinfo {author}
  {\bibfnamefont {A.}~\bibnamefont {{Waczynski}}}, \bibinfo {author}
  {\bibfnamefont {Y.}~\bibnamefont {{Wen}}}, and\ \bibinfo {author}
  {\bibnamefont {{WFC3 Team}}},\ }in\ \href@noop {} {\emph {\bibinfo
  {booktitle} {American Astronomical Society Meeting Abstracts}}},\ \bibinfo
  {series} {American Astronomical Society Meeting Abstracts}, Vol.\ \bibinfo
  {volume} {205}\ (\bibinfo {year} {2004})\ p.\ \bibinfo {pages}
  {05.02}\BibitemShut {NoStop}%
\bibitem [{\citenamefont {{Kidder}}(1995)}]{Kidder}%
  \BibitemOpen
  \bibfield  {author} {\bibinfo {author} {\bibfnamefont {L.~E.}\ \bibnamefont
  {{Kidder}}},\ }\href {https://doi.org/10.1103/PhysRevD.52.821} {\bibfield
  {journal} {\bibinfo  {journal} {\prd}\ }\textbf {\bibinfo {volume} {52}},\
  \bibinfo {pages} {821} (\bibinfo {year} {1995})},\ \Eprint
  {https://arxiv.org/abs/gr-qc/9506022} {arXiv:gr-qc/9506022 [gr-qc]}
  \BibitemShut {NoStop}%
\bibitem [{\citenamefont {{Lindegren}}\ \emph {et~al.}(2020)\citenamefont
  {{Lindegren}}, \citenamefont {{Klioner}}, \citenamefont {{Hern{\'a}ndez}},
  \citenamefont {{Bombrun}}, \citenamefont {{Ramos-Lerate}}, \citenamefont
  {{Steidelm{\"u}ller}}, \citenamefont {{Bastian}}, \citenamefont {{Biermann}}
  \emph {et~al.}}]{eDR3}%
  \BibitemOpen
  \bibfield  {author} {\bibinfo {author} {\bibfnamefont {L.}~\bibnamefont
  {{Lindegren}}}, \bibinfo {author} {\bibfnamefont {S.~A.}\ \bibnamefont
  {{Klioner}}}, \bibinfo {author} {\bibfnamefont {J.}~\bibnamefont
  {{Hern{\'a}ndez}}}, \bibinfo {author} {\bibfnamefont {A.}~\bibnamefont
  {{Bombrun}}}, \bibinfo {author} {\bibfnamefont {M.}~\bibnamefont
  {{Ramos-Lerate}}}, \bibinfo {author} {\bibfnamefont {H.}~\bibnamefont
  {{Steidelm{\"u}ller}}}, \bibinfo {author} {\bibfnamefont {U.}~\bibnamefont
  {{Bastian}}}, \bibinfo {author} {\bibfnamefont {M.}~\bibnamefont
  {{Biermann}}},  \emph {et~al.},\ }\href@noop {} {\bibfield  {journal}
  {\bibinfo  {journal} {arXiv e-prints}\ ,\ \bibinfo {eid} {arXiv:2012.03380}}
  (\bibinfo {year} {2020})},\ \Eprint {https://arxiv.org/abs/2012.03380}
  {arXiv:2012.03380 [astro-ph.IM]} \BibitemShut {NoStop}%
\bibitem [{\citenamefont {{Luri}}\ \emph {et~al.}(2018)\citenamefont {{Luri}},
  \citenamefont {{Brown}}, \citenamefont {{Sarro}}, \citenamefont {{Arenou}},
  \citenamefont {{Bailer-Jones}}, \citenamefont {{Castro-Ginard}},
  \citenamefont {{de Bruijne}}, \citenamefont {{Prusti}}, \citenamefont
  {{Babusiaux}},\ and\ \citenamefont {{Delgado}}}]{Luri2018}%
  \BibitemOpen
  \bibfield  {author} {\bibinfo {author} {\bibfnamefont {X.}~\bibnamefont
  {{Luri}}}, \bibinfo {author} {\bibfnamefont {A.~G.~A.}\ \bibnamefont
  {{Brown}}}, \bibinfo {author} {\bibfnamefont {L.~M.}\ \bibnamefont
  {{Sarro}}}, \bibinfo {author} {\bibfnamefont {F.}~\bibnamefont {{Arenou}}},
  \bibinfo {author} {\bibfnamefont {C.~A.~L.}\ \bibnamefont {{Bailer-Jones}}},
  \bibinfo {author} {\bibfnamefont {A.}~\bibnamefont {{Castro-Ginard}}},
  \bibinfo {author} {\bibfnamefont {J.}~\bibnamefont {{de Bruijne}}}, \bibinfo
  {author} {\bibfnamefont {T.}~\bibnamefont {{Prusti}}}, \bibinfo {author}
  {\bibfnamefont {C.}~\bibnamefont {{Babusiaux}}}, and\ \bibinfo {author}
  {\bibfnamefont {H.~E.}\ \bibnamefont {{Delgado}}},\ }\href
  {https://doi.org/10.1051/0004-6361/201832964} {\bibfield  {journal} {\bibinfo
   {journal} {\aap}\ }\textbf {\bibinfo {volume} {616}},\ \bibinfo {eid} {A9}
  (\bibinfo {year} {2018})},\ \Eprint {https://arxiv.org/abs/1804.09376}
  {arXiv:1804.09376 [astro-ph.IM]} \BibitemShut {NoStop}%
\bibitem [{\citenamefont {{Thrane}}\ and\ \citenamefont
  {{Romano}}(2013)}]{Thrane2013}%
  \BibitemOpen
  \bibfield  {author} {\bibinfo {author} {\bibfnamefont {E.}~\bibnamefont
  {{Thrane}}}and\ \bibinfo {author} {\bibfnamefont {J.~D.}\ \bibnamefont
  {{Romano}}},\ }\href {https://doi.org/10.1103/PhysRevD.88.124032} {\bibfield
  {journal} {\bibinfo  {journal} {\prd}\ }\textbf {\bibinfo {volume} {88}},\
  \bibinfo {eid} {124032} (\bibinfo {year} {2013})},\ \Eprint
  {https://arxiv.org/abs/1310.5300} {arXiv:1310.5300 [astro-ph.IM]}
  \BibitemShut {NoStop}%
\bibitem [{\citenamefont {{Robson}}\ \emph {et~al.}(2019)\citenamefont
  {{Robson}}, \citenamefont {{Cornish}},\ and\ \citenamefont
  {{Liu}}}]{Robson2019}%
  \BibitemOpen
  \bibfield  {author} {\bibinfo {author} {\bibfnamefont {T.}~\bibnamefont
  {{Robson}}}, \bibinfo {author} {\bibfnamefont {N.~J.}\ \bibnamefont
  {{Cornish}}}, and\ \bibinfo {author} {\bibfnamefont {C.}~\bibnamefont
  {{Liu}}},\ }\href {https://doi.org/10.1088/1361-6382/ab1101} {\bibfield
  {journal} {\bibinfo  {journal} {Classical and Quantum Gravity}\ }\textbf
  {\bibinfo {volume} {36}},\ \bibinfo {eid} {105011} (\bibinfo {year}
  {2019})},\ \Eprint {https://arxiv.org/abs/1803.01944} {arXiv:1803.01944
  [astro-ph.HE]} \BibitemShut {NoStop}%
\bibitem [{\citenamefont {{Cutler}}\ and\ \citenamefont
  {{Flanagan}}(1994)}]{Cutler1994}%
  \BibitemOpen
  \bibfield  {author} {\bibinfo {author} {\bibfnamefont {C.}~\bibnamefont
  {{Cutler}}}and\ \bibinfo {author} {\bibfnamefont {{\'E}.~E.}\ \bibnamefont
  {{Flanagan}}},\ }\href {https://doi.org/10.1103/PhysRevD.49.2658} {\bibfield
  {journal} {\bibinfo  {journal} {\prd}\ }\textbf {\bibinfo {volume} {49}},\
  \bibinfo {pages} {2658} (\bibinfo {year} {1994})},\ \Eprint
  {https://arxiv.org/abs/gr-qc/9402014} {arXiv:gr-qc/9402014 [gr-qc]}
  \BibitemShut {NoStop}%
\bibitem [{\citenamefont {{Moore}}\ \emph
  {et~al.}(2015{\natexlab{b}})\citenamefont {{Moore}}, \citenamefont {{Cole}},\
  and\ \citenamefont {{Berry}}}]{GWPlotter}%
  \BibitemOpen
  \bibfield  {author} {\bibinfo {author} {\bibfnamefont {C.~J.}\ \bibnamefont
  {{Moore}}}, \bibinfo {author} {\bibfnamefont {R.~H.}\ \bibnamefont {{Cole}}},
  and\ \bibinfo {author} {\bibfnamefont {C.~P.~L.}\ \bibnamefont {{Berry}}},\
  }\href {https://doi.org/10.1088/0264-9381/32/1/015014} {\bibfield  {journal}
  {\bibinfo  {journal} {Classical and Quantum Gravity}\ }\textbf {\bibinfo
  {volume} {32}},\ \bibinfo {eid} {015014} (\bibinfo {year}
  {2015}{\natexlab{b}})},\ \Eprint {https://arxiv.org/abs/1408.0740}
  {arXiv:1408.0740 [gr-qc]} \BibitemShut {NoStop}%
\bibitem [{\citenamefont {{Sesana}}\ \emph {et~al.}(2004)\citenamefont
  {{Sesana}}, \citenamefont {{Haardt}}, \citenamefont {{Madau}},\ and\
  \citenamefont {{Volonteri}}}]{Sesana2004}%
  \BibitemOpen
  \bibfield  {author} {\bibinfo {author} {\bibfnamefont {A.}~\bibnamefont
  {{Sesana}}}, \bibinfo {author} {\bibfnamefont {F.}~\bibnamefont {{Haardt}}},
  \bibinfo {author} {\bibfnamefont {P.}~\bibnamefont {{Madau}}}, and\ \bibinfo
  {author} {\bibfnamefont {M.}~\bibnamefont {{Volonteri}}},\ }\href
  {https://doi.org/10.1086/422185} {\bibfield  {journal} {\bibinfo  {journal}
  {\apj}\ }\textbf {\bibinfo {volume} {611}},\ \bibinfo {pages} {623} (\bibinfo
  {year} {2004})},\ \Eprint {https://arxiv.org/abs/astro-ph/0401543}
  {arXiv:astro-ph/0401543 [astro-ph]} \BibitemShut {NoStop}%
\bibitem [{\citenamefont {{Clarkson}}\ \emph {et~al.}(2008)\citenamefont
  {{Clarkson}}, \citenamefont {{Sahu}}, \citenamefont {{Anderson}},
  \citenamefont {{Smith}}, \citenamefont {{Brown}}, \citenamefont {{Rich}},
  \citenamefont {{Casertano}}, \citenamefont {{Bond}}, \citenamefont {{Livio}},
  \citenamefont {{Minniti}}, \citenamefont {{Panagia}}, \citenamefont
  {{Renzini}}, \citenamefont {{Valenti}},\ and\ \citenamefont
  {{Zoccali}}}]{Clarkson2008}%
  \BibitemOpen
  \bibfield  {author} {\bibinfo {author} {\bibfnamefont {W.}~\bibnamefont
  {{Clarkson}}}, \bibinfo {author} {\bibfnamefont {K.}~\bibnamefont {{Sahu}}},
  \bibinfo {author} {\bibfnamefont {J.}~\bibnamefont {{Anderson}}}, \bibinfo
  {author} {\bibfnamefont {T.~E.}\ \bibnamefont {{Smith}}}, \bibinfo {author}
  {\bibfnamefont {T.~M.}\ \bibnamefont {{Brown}}}, \bibinfo {author}
  {\bibfnamefont {R.~M.}\ \bibnamefont {{Rich}}}, \bibinfo {author}
  {\bibfnamefont {S.}~\bibnamefont {{Casertano}}}, \bibinfo {author}
  {\bibfnamefont {H.~E.}\ \bibnamefont {{Bond}}}, \bibinfo {author}
  {\bibfnamefont {M.}~\bibnamefont {{Livio}}}, \bibinfo {author} {\bibfnamefont
  {D.}~\bibnamefont {{Minniti}}}, \bibinfo {author} {\bibfnamefont
  {N.}~\bibnamefont {{Panagia}}}, \bibinfo {author} {\bibfnamefont
  {A.}~\bibnamefont {{Renzini}}}, \bibinfo {author} {\bibfnamefont
  {J.}~\bibnamefont {{Valenti}}}, and\ \bibinfo {author} {\bibfnamefont
  {M.}~\bibnamefont {{Zoccali}}},\ }\href {https://doi.org/10.1086/590378}
  {\bibfield  {journal} {\bibinfo  {journal} {\apj}\ }\textbf {\bibinfo
  {volume} {684}},\ \bibinfo {pages} {1110} (\bibinfo {year} {2008})},\ \Eprint
  {https://arxiv.org/abs/0809.1682} {arXiv:0809.1682 [astro-ph]} \BibitemShut
  {NoStop}%
\bibitem [{\citenamefont {{Lindegren}}\ and\ \citenamefont
  {{Bastian}}(2010)}]{Lindegren}%
  \BibitemOpen
  \bibfield  {author} {\bibinfo {author} {\bibfnamefont {L.}~\bibnamefont
  {{Lindegren}}}and\ \bibinfo {author} {\bibfnamefont {U.}~\bibnamefont
  {{Bastian}}},\ }in\ \href {https://doi.org/10.1051/eas/1045018} {\emph
  {\bibinfo {booktitle} {EAS Publications Series}}},\ \bibinfo {series} {EAS
  Publications Series}, Vol.~\bibinfo {volume} {45}\ (\bibinfo {year} {2010})\
  pp.\ \bibinfo {pages} {109--114}\BibitemShut {NoStop}%
\bibitem [{\citenamefont {{van Haasteren}}\ \emph {et~al.}(2011)\citenamefont
  {{van Haasteren}}, \citenamefont {{Levin}}, \citenamefont {{Janssen}},
  \citenamefont {{Lazaridis}}, \citenamefont {{Kramer}}, \citenamefont
  {{Stappers}}, \citenamefont {{Desvignes}}, \citenamefont {{Purver}},
  \citenamefont {{Lyne}}, \citenamefont {{Ferdman}}, \citenamefont {{Jessner}},
  \citenamefont {{Cognard}}, \citenamefont {{Theureau}}, \citenamefont
  {{D'Amico}}, \citenamefont {{Possenti}}, \citenamefont {{Burgay}},
  \citenamefont {{Corongiu}}, \citenamefont {{Hessels}}, \citenamefont
  {{Smits}},\ and\ \citenamefont {{Verbiest}}}]{vanHaasteren2011}%
  \BibitemOpen
  \bibfield  {author} {\bibinfo {author} {\bibfnamefont {R.}~\bibnamefont {{van
  Haasteren}}}, \bibinfo {author} {\bibfnamefont {Y.}~\bibnamefont {{Levin}}},
  \bibinfo {author} {\bibfnamefont {G.~H.}\ \bibnamefont {{Janssen}}}, \bibinfo
  {author} {\bibfnamefont {K.}~\bibnamefont {{Lazaridis}}}, \bibinfo {author}
  {\bibfnamefont {M.}~\bibnamefont {{Kramer}}}, \bibinfo {author}
  {\bibfnamefont {B.~W.}\ \bibnamefont {{Stappers}}}, \bibinfo {author}
  {\bibfnamefont {G.}~\bibnamefont {{Desvignes}}}, \bibinfo {author}
  {\bibfnamefont {M.~B.}\ \bibnamefont {{Purver}}}, \bibinfo {author}
  {\bibfnamefont {A.~G.}\ \bibnamefont {{Lyne}}}, \bibinfo {author}
  {\bibfnamefont {R.~D.}\ \bibnamefont {{Ferdman}}}, \bibinfo {author}
  {\bibfnamefont {A.}~\bibnamefont {{Jessner}}}, \bibinfo {author}
  {\bibfnamefont {I.}~\bibnamefont {{Cognard}}}, \bibinfo {author}
  {\bibfnamefont {G.}~\bibnamefont {{Theureau}}}, \bibinfo {author}
  {\bibfnamefont {N.}~\bibnamefont {{D'Amico}}}, \bibinfo {author}
  {\bibfnamefont {A.}~\bibnamefont {{Possenti}}}, \bibinfo {author}
  {\bibfnamefont {M.}~\bibnamefont {{Burgay}}}, \bibinfo {author}
  {\bibfnamefont {A.}~\bibnamefont {{Corongiu}}}, \bibinfo {author}
  {\bibfnamefont {J.~W.~T.}\ \bibnamefont {{Hessels}}}, \bibinfo {author}
  {\bibfnamefont {R.}~\bibnamefont {{Smits}}}, and\ \bibinfo {author}
  {\bibfnamefont {J.~P.~W.}\ \bibnamefont {{Verbiest}}},\ }\href
  {https://doi.org/10.1111/j.1365-2966.2011.18613.x} {\bibfield  {journal}
  {\bibinfo  {journal} {\mnras}\ }\textbf {\bibinfo {volume} {414}},\ \bibinfo
  {pages} {3117} (\bibinfo {year} {2011})},\ \Eprint
  {https://arxiv.org/abs/1103.0576} {arXiv:1103.0576 [astro-ph.CO]}
  \BibitemShut {NoStop}%
\bibitem [{\citenamefont {{Arzoumanian}}\ \emph {et~al.}(2018)\citenamefont
  {{Arzoumanian}}, \citenamefont {{Baker}}, \citenamefont {{Brazier}},
  \citenamefont {{Burke-Spolaor}}, \citenamefont {{Chamberlin}}, \citenamefont
  {{Chatterjee}}, \citenamefont {{Christy}}, \citenamefont {{Cordes}},
  \citenamefont {{Cornish}}, \citenamefont {{Crawford}}, \citenamefont
  {{Thankful Cromartie}}, \citenamefont {{Crowter}}, \citenamefont {{DeCesar}},
  \citenamefont {{Demorest}}, \citenamefont {{Dolch}}, \citenamefont {{Ellis}},
  \citenamefont {others},\ and\ \citenamefont {{NANOGrav
  Collaboration}}}]{NANOGrav112018}%
  \BibitemOpen
  \bibfield  {author} {\bibinfo {author} {\bibfnamefont {Z.}~\bibnamefont
  {{Arzoumanian}}}, \bibinfo {author} {\bibfnamefont {P.~T.}\ \bibnamefont
  {{Baker}}}, \bibinfo {author} {\bibfnamefont {A.}~\bibnamefont {{Brazier}}},
  \bibinfo {author} {\bibfnamefont {S.}~\bibnamefont {{Burke-Spolaor}}},
  \bibinfo {author} {\bibfnamefont {S.~J.}\ \bibnamefont {{Chamberlin}}},
  \bibinfo {author} {\bibfnamefont {S.}~\bibnamefont {{Chatterjee}}}, \bibinfo
  {author} {\bibfnamefont {B.}~\bibnamefont {{Christy}}}, \bibinfo {author}
  {\bibfnamefont {J.~M.}\ \bibnamefont {{Cordes}}}, \bibinfo {author}
  {\bibfnamefont {N.~J.}\ \bibnamefont {{Cornish}}}, \bibinfo {author}
  {\bibfnamefont {F.}~\bibnamefont {{Crawford}}}, \bibinfo {author}
  {\bibfnamefont {H.}~\bibnamefont {{Thankful Cromartie}}}, \bibinfo {author}
  {\bibfnamefont {K.}~\bibnamefont {{Crowter}}}, \bibinfo {author}
  {\bibfnamefont {M.}~\bibnamefont {{DeCesar}}}, \bibinfo {author}
  {\bibfnamefont {P.~B.}\ \bibnamefont {{Demorest}}}, \bibinfo {author}
  {\bibfnamefont {T.}~\bibnamefont {{Dolch}}}, \bibinfo {author} {\bibfnamefont
  {J.~A.}\ \bibnamefont {{Ellis}}}, \bibinfo {author} {\bibnamefont {others}},
  and\ \bibinfo {author} {\bibnamefont {{NANOGrav Collaboration}}},\ }\href
  {https://doi.org/10.3847/1538-4357/aabd3b} {\bibfield  {journal} {\bibinfo
  {journal} {\apj}\ }\textbf {\bibinfo {volume} {859}},\ \bibinfo {eid} {47}
  (\bibinfo {year} {2018})},\ \Eprint {https://arxiv.org/abs/1801.02617}
  {arXiv:1801.02617 [astro-ph.HE]} \BibitemShut {NoStop}%
\bibitem [{\citenamefont {{Mingarelli}}\ \emph {et~al.}(2017)\citenamefont
  {{Mingarelli}}, \citenamefont {{Lazio}}, \citenamefont {{Sesana}},
  \citenamefont {{Greene}}, \citenamefont {{Ellis}}, \citenamefont {{Ma}},
  \citenamefont {{Croft}}, \citenamefont {{Burke-Spolaor}},\ and\ \citenamefont
  {{Taylor}}}]{Mingarelli2017}%
  \BibitemOpen
  \bibfield  {author} {\bibinfo {author} {\bibfnamefont {C.~M.~F.}\
  \bibnamefont {{Mingarelli}}}, \bibinfo {author} {\bibfnamefont {T.~J.~W.}\
  \bibnamefont {{Lazio}}}, \bibinfo {author} {\bibfnamefont {A.}~\bibnamefont
  {{Sesana}}}, \bibinfo {author} {\bibfnamefont {J.~E.}\ \bibnamefont
  {{Greene}}}, \bibinfo {author} {\bibfnamefont {J.~A.}\ \bibnamefont
  {{Ellis}}}, \bibinfo {author} {\bibfnamefont {C.-P.}\ \bibnamefont {{Ma}}},
  \bibinfo {author} {\bibfnamefont {S.}~\bibnamefont {{Croft}}}, \bibinfo
  {author} {\bibfnamefont {S.}~\bibnamefont {{Burke-Spolaor}}}, and\ \bibinfo
  {author} {\bibfnamefont {S.~R.}\ \bibnamefont {{Taylor}}},\ }\href
  {https://doi.org/10.1038/s41550-017-0299-6} {\bibfield  {journal} {\bibinfo
  {journal} {Nature Astronomy}\ }\textbf {\bibinfo {volume} {1}},\ \bibinfo
  {pages} {886} (\bibinfo {year} {2017})},\ \Eprint
  {https://arxiv.org/abs/1708.03491} {arXiv:1708.03491 [astro-ph.GA]}
  \BibitemShut {NoStop}%
\bibitem [{\citenamefont {Kruk}()}]{JPL}%
  \BibitemOpen
  \bibfield  {author} {\bibinfo {author} {\bibfnamefont {J.}~\bibnamefont
  {Kruk}},\ }\href@noop {} {}\bibinfo {howpublished} {personal
  communication}\BibitemShut {NoStop}%
\bibitem [{\citenamefont {{Weltman}}\ \emph {et~al.}(2020)\citenamefont
  {{Weltman}}, \citenamefont {{Bull}}, \citenamefont {{Camera}}, \citenamefont
  {{Kelley}}, \citenamefont {{Padmanabhan}}, \citenamefont {{Pritchard}},
  \citenamefont {{Raccanelli}}, \citenamefont {{Riemer-S{\o}rensen}},
  \citenamefont {{Shao}}, \citenamefont {{Andrianomena}}, \citenamefont
  {{Athanassoula}}, \citenamefont {{Bacon}}, \citenamefont {{Barkana}},
  \citenamefont {{Bertone}}, \citenamefont {{B{\oe}hm}}, \citenamefont
  {{Bonvin}}, \citenamefont {{Bosma}}, \citenamefont {{Br{\"u}ggen}} \emph
  {et~al.}}]{Weltman2020}%
  \BibitemOpen
  \bibfield  {author} {\bibinfo {author} {\bibfnamefont {A.}~\bibnamefont
  {{Weltman}}}, \bibinfo {author} {\bibfnamefont {P.}~\bibnamefont {{Bull}}},
  \bibinfo {author} {\bibfnamefont {S.}~\bibnamefont {{Camera}}}, \bibinfo
  {author} {\bibfnamefont {K.}~\bibnamefont {{Kelley}}}, \bibinfo {author}
  {\bibfnamefont {H.}~\bibnamefont {{Padmanabhan}}}, \bibinfo {author}
  {\bibfnamefont {J.}~\bibnamefont {{Pritchard}}}, \bibinfo {author}
  {\bibfnamefont {A.}~\bibnamefont {{Raccanelli}}}, \bibinfo {author}
  {\bibfnamefont {S.}~\bibnamefont {{Riemer-S{\o}rensen}}}, \bibinfo {author}
  {\bibfnamefont {L.}~\bibnamefont {{Shao}}}, \bibinfo {author} {\bibfnamefont
  {S.}~\bibnamefont {{Andrianomena}}}, \bibinfo {author} {\bibfnamefont
  {E.}~\bibnamefont {{Athanassoula}}}, \bibinfo {author} {\bibfnamefont
  {D.}~\bibnamefont {{Bacon}}}, \bibinfo {author} {\bibfnamefont
  {R.}~\bibnamefont {{Barkana}}}, \bibinfo {author} {\bibfnamefont
  {G.}~\bibnamefont {{Bertone}}}, \bibinfo {author} {\bibfnamefont
  {C.}~\bibnamefont {{B{\oe}hm}}}, \bibinfo {author} {\bibfnamefont
  {C.}~\bibnamefont {{Bonvin}}}, \bibinfo {author} {\bibfnamefont
  {A.}~\bibnamefont {{Bosma}}}, \bibinfo {author} {\bibfnamefont
  {M.}~\bibnamefont {{Br{\"u}ggen}}},  \emph {et~al.},\ }\href
  {https://doi.org/10.1017/pasa.2019.42} {\bibfield  {journal} {\bibinfo
  {journal} {\pasa}\ }\textbf {\bibinfo {volume} {37}},\ \bibinfo {eid} {e002}
  (\bibinfo {year} {2020})},\ \Eprint {https://arxiv.org/abs/1810.02680}
  {arXiv:1810.02680 [astro-ph.CO]} \BibitemShut {NoStop}%
\bibitem [{SKA(2018)}]{SKAinfo}%
  \BibitemOpen
  \href
  {https://www.skatelescope.org/wp-content/uploads/2018/08/16231-factsheet-telescopes-v71.pdf}
  {\bibinfo {title} {{Technical Information - the Telescopes}}} (\bibinfo
  {year} {2018})\BibitemShut {NoStop}%
\bibitem [{SKA(2015)}]{SKAreq}%
  \BibitemOpen
  \href
  {https://astronomers.skatelescope.org/wp-content/uploads/2015/11/SKA-TEL-SKO-0000007_SKA1_Level_0_Science_RequirementsRev02-part-1-signed.pdf}
  {\emph {\bibinfo {title} {{SKA1 level 0 science requirements}}}},\ \bibinfo
  {type} {Tech. Rep.}\ \bibinfo {number} {SKA-TEL-SKO-0000007}\ (\bibinfo
  {year} {2015})\BibitemShut {NoStop}%
\bibitem [{\citenamefont {{Mellier}}(2012)}]{Euclid2012}%
  \BibitemOpen
  \bibfield  {author} {\bibinfo {author} {\bibfnamefont {Y.}~\bibnamefont
  {{Mellier}}},\ }in\ \href@noop {} {\emph {\bibinfo {booktitle} {Science from
  the Next Generation Imaging and Spectroscopic Surveys}}}\ (\bibinfo {year}
  {2012})\ p.~\bibinfo {pages} {3}\BibitemShut {NoStop}%
\bibitem [{\citenamefont {{Ivezi{\'c}}}\ \emph {et~al.}(2019)\citenamefont
  {{Ivezi{\'c}}}, \citenamefont {{Kahn}}, \citenamefont {{Tyson}},
  \citenamefont {{Abel}}, \citenamefont {{Acosta}}, \citenamefont {{Allsman}},
  \citenamefont {{Alonso}}, \citenamefont {{AlSayyad}}, \citenamefont
  {{Anderson}}, \citenamefont {{Andrew}}, \citenamefont {{Angel}},
  \citenamefont {{Angeli}}, \citenamefont {{Ansari}}, \citenamefont
  {{Antilogus}}, \citenamefont {{Araujo}}, \citenamefont {{Armstrong}},
  \citenamefont {{Arndt}}, \citenamefont {{Astier}} \emph {et~al.}}]{LSST2019}%
  \BibitemOpen
  \bibfield  {author} {\bibinfo {author} {\bibfnamefont {{\v{Z}}.}~\bibnamefont
  {{Ivezi{\'c}}}}, \bibinfo {author} {\bibfnamefont {S.~M.}\ \bibnamefont
  {{Kahn}}}, \bibinfo {author} {\bibfnamefont {J.~A.}\ \bibnamefont {{Tyson}}},
  \bibinfo {author} {\bibfnamefont {B.}~\bibnamefont {{Abel}}}, \bibinfo
  {author} {\bibfnamefont {E.}~\bibnamefont {{Acosta}}}, \bibinfo {author}
  {\bibfnamefont {R.}~\bibnamefont {{Allsman}}}, \bibinfo {author}
  {\bibfnamefont {D.}~\bibnamefont {{Alonso}}}, \bibinfo {author}
  {\bibfnamefont {Y.}~\bibnamefont {{AlSayyad}}}, \bibinfo {author}
  {\bibfnamefont {S.~F.}\ \bibnamefont {{Anderson}}}, \bibinfo {author}
  {\bibfnamefont {J.}~\bibnamefont {{Andrew}}}, \bibinfo {author}
  {\bibfnamefont {J.~R.~P.}\ \bibnamefont {{Angel}}}, \bibinfo {author}
  {\bibfnamefont {G.~Z.}\ \bibnamefont {{Angeli}}}, \bibinfo {author}
  {\bibfnamefont {R.}~\bibnamefont {{Ansari}}}, \bibinfo {author}
  {\bibfnamefont {P.}~\bibnamefont {{Antilogus}}}, \bibinfo {author}
  {\bibfnamefont {C.}~\bibnamefont {{Araujo}}}, \bibinfo {author}
  {\bibfnamefont {R.}~\bibnamefont {{Armstrong}}}, \bibinfo {author}
  {\bibfnamefont {K.~T.}\ \bibnamefont {{Arndt}}}, \bibinfo {author}
  {\bibfnamefont {P.}~\bibnamefont {{Astier}}},  \emph {et~al.},\ }\href
  {https://doi.org/10.3847/1538-4357/ab042c} {\bibfield  {journal} {\bibinfo
  {journal} {\apj}\ }\textbf {\bibinfo {volume} {873}},\ \bibinfo {eid} {111}
  (\bibinfo {year} {2019})},\ \Eprint {https://arxiv.org/abs/0805.2366}
  {arXiv:0805.2366 [astro-ph]} \BibitemShut {NoStop}%
\bibitem [{\citenamefont {{McKinnon}}\ \emph {et~al.}(2019)\citenamefont
  {{McKinnon}}, \citenamefont {{Beasley}}, \citenamefont {{Murphy}},
  \citenamefont {{Selina}}, \citenamefont {{Farnsworth}},\ and\ \citenamefont
  {{Walter}}}]{ngVLA2019}%
  \BibitemOpen
  \bibfield  {author} {\bibinfo {author} {\bibfnamefont {M.}~\bibnamefont
  {{McKinnon}}}, \bibinfo {author} {\bibfnamefont {A.}~\bibnamefont
  {{Beasley}}}, \bibinfo {author} {\bibfnamefont {E.}~\bibnamefont {{Murphy}}},
  \bibinfo {author} {\bibfnamefont {R.}~\bibnamefont {{Selina}}}, \bibinfo
  {author} {\bibfnamefont {R.}~\bibnamefont {{Farnsworth}}}, and\ \bibinfo
  {author} {\bibfnamefont {A.}~\bibnamefont {{Walter}}},\ }in\ \href@noop {}
  {\emph {\bibinfo {booktitle} {\baas}}},\ Vol.~\bibinfo {volume} {51}\
  (\bibinfo {year} {2019})\ p.~\bibinfo {pages} {81}\BibitemShut {NoStop}%
\bibitem [{\citenamefont {{Jaffe}}(2004)}]{Jaffe2004}%
  \BibitemOpen
  \bibfield  {author} {\bibinfo {author} {\bibfnamefont {A.~H.}\ \bibnamefont
  {{Jaffe}}},\ }\href {https://doi.org/10.1016/j.newar.2004.09.018} {\bibfield
  {journal} {\bibinfo  {journal} {\nar}\ }\textbf {\bibinfo {volume} {48}},\
  \bibinfo {pages} {1483} (\bibinfo {year} {2004})},\ \Eprint
  {https://arxiv.org/abs/astro-ph/0409637} {arXiv:astro-ph/0409637 [astro-ph]}
  \BibitemShut {NoStop}%
\bibitem [{\citenamefont {{Percival}}\ \emph {et~al.}(2019)\citenamefont
  {{Percival}}, \citenamefont {{Balogh}}, \citenamefont {{Bond}}, \citenamefont
  {{Bovy}}, \citenamefont {{Carlberg}}, \citenamefont {{Chapman}},
  \citenamefont {{Cote}}, \citenamefont {{Cowan}}, \citenamefont {{Fabbro}},
  \citenamefont {{Ferrarese}}, \citenamefont {{Gwyn}}, \citenamefont
  {{Hlozek}}, \citenamefont {{Hudson}}, \citenamefont {{Hutchings}},
  \citenamefont {{Kavelaars}}, \citenamefont {{Lang}}, \citenamefont
  {{McConnachie}}, \citenamefont {{Muzzin}}, \citenamefont {{Parker}},
  \citenamefont {{Pritchet}}, \citenamefont {{Sawicki}}, \citenamefont
  {{Schade}}, \citenamefont {{Scott}}, \citenamefont {{Smith}}, \citenamefont
  {{Spekkens}}, \citenamefont {{Taylor}},\ and\ \citenamefont
  {{Willott}}}]{Euclid2019}%
  \BibitemOpen
  \bibfield  {author} {\bibinfo {author} {\bibfnamefont {W.}~\bibnamefont
  {{Percival}}}, \bibinfo {author} {\bibfnamefont {M.}~\bibnamefont
  {{Balogh}}}, \bibinfo {author} {\bibfnamefont {D.}~\bibnamefont {{Bond}}},
  \bibinfo {author} {\bibfnamefont {J.}~\bibnamefont {{Bovy}}}, \bibinfo
  {author} {\bibfnamefont {R.}~\bibnamefont {{Carlberg}}}, \bibinfo {author}
  {\bibfnamefont {S.}~\bibnamefont {{Chapman}}}, \bibinfo {author}
  {\bibfnamefont {P.}~\bibnamefont {{Cote}}}, \bibinfo {author} {\bibfnamefont
  {N.}~\bibnamefont {{Cowan}}}, \bibinfo {author} {\bibfnamefont
  {S.}~\bibnamefont {{Fabbro}}}, \bibinfo {author} {\bibfnamefont
  {L.}~\bibnamefont {{Ferrarese}}}, \bibinfo {author} {\bibfnamefont
  {S.}~\bibnamefont {{Gwyn}}}, \bibinfo {author} {\bibfnamefont
  {R.}~\bibnamefont {{Hlozek}}}, \bibinfo {author} {\bibfnamefont
  {M.}~\bibnamefont {{Hudson}}}, \bibinfo {author} {\bibfnamefont
  {J.}~\bibnamefont {{Hutchings}}}, \bibinfo {author} {\bibfnamefont
  {J.}~\bibnamefont {{Kavelaars}}}, \bibinfo {author} {\bibfnamefont
  {D.}~\bibnamefont {{Lang}}}, \bibinfo {author} {\bibfnamefont
  {A.}~\bibnamefont {{McConnachie}}}, \bibinfo {author} {\bibfnamefont
  {A.}~\bibnamefont {{Muzzin}}}, \bibinfo {author} {\bibfnamefont
  {L.}~\bibnamefont {{Parker}}}, \bibinfo {author} {\bibfnamefont
  {C.}~\bibnamefont {{Pritchet}}}, \bibinfo {author} {\bibfnamefont
  {M.}~\bibnamefont {{Sawicki}}}, \bibinfo {author} {\bibfnamefont
  {D.}~\bibnamefont {{Schade}}}, \bibinfo {author} {\bibfnamefont
  {D.}~\bibnamefont {{Scott}}}, \bibinfo {author} {\bibfnamefont
  {K.}~\bibnamefont {{Smith}}}, \bibinfo {author} {\bibfnamefont
  {K.}~\bibnamefont {{Spekkens}}}, \bibinfo {author} {\bibfnamefont
  {J.}~\bibnamefont {{Taylor}}}, and\ \bibinfo {author} {\bibfnamefont
  {C.}~\bibnamefont {{Willott}}},\ }in\ \href
  {https://doi.org/10.5281/zenodo.3758532} {\emph {\bibinfo {booktitle}
  {Canadian Long Range Plan for Astronony and Astrophysics White Papers}}},\
  Vol.\ \bibinfo {volume} {2020}\ (\bibinfo {year} {2019})\ p.~\bibinfo {pages}
  {20}\BibitemShut {NoStop}%
\bibitem [{\citenamefont {{Gaudi}}\ \emph {et~al.}(2020)\citenamefont
  {{Gaudi}}, \citenamefont {{Seager}}, \citenamefont {{Mennesson}},
  \citenamefont {{Kiessling}}, \citenamefont {{Warfield}}, \citenamefont
  {{Cahoy}}, \citenamefont {{Clarke}}, \citenamefont {{Domagal-Goldman}},
  \citenamefont {{Feinberg}}, \citenamefont {{Guyon}}, \citenamefont
  {{Kasdin}}, \citenamefont {{Mawet}}, \citenamefont {{Plavchan}},
  \citenamefont {{Robinson}}, \citenamefont {{Rogers}}, \citenamefont
  {{Scowen}} \emph {et~al.}}]{Habex2020}%
  \BibitemOpen
  \bibfield  {author} {\bibinfo {author} {\bibfnamefont {B.~S.}\ \bibnamefont
  {{Gaudi}}}, \bibinfo {author} {\bibfnamefont {S.}~\bibnamefont {{Seager}}},
  \bibinfo {author} {\bibfnamefont {B.}~\bibnamefont {{Mennesson}}}, \bibinfo
  {author} {\bibfnamefont {A.}~\bibnamefont {{Kiessling}}}, \bibinfo {author}
  {\bibfnamefont {K.}~\bibnamefont {{Warfield}}}, \bibinfo {author}
  {\bibfnamefont {K.}~\bibnamefont {{Cahoy}}}, \bibinfo {author} {\bibfnamefont
  {J.~T.}\ \bibnamefont {{Clarke}}}, \bibinfo {author} {\bibfnamefont
  {S.}~\bibnamefont {{Domagal-Goldman}}}, \bibinfo {author} {\bibfnamefont
  {L.}~\bibnamefont {{Feinberg}}}, \bibinfo {author} {\bibfnamefont
  {O.}~\bibnamefont {{Guyon}}}, \bibinfo {author} {\bibfnamefont
  {J.}~\bibnamefont {{Kasdin}}}, \bibinfo {author} {\bibfnamefont
  {D.}~\bibnamefont {{Mawet}}}, \bibinfo {author} {\bibfnamefont
  {P.}~\bibnamefont {{Plavchan}}}, \bibinfo {author} {\bibfnamefont
  {T.}~\bibnamefont {{Robinson}}}, \bibinfo {author} {\bibfnamefont
  {L.}~\bibnamefont {{Rogers}}}, \bibinfo {author} {\bibfnamefont
  {P.}~\bibnamefont {{Scowen}}},  \emph {et~al.},\ }\href@noop {} {\bibfield
  {journal} {\bibinfo  {journal} {arXiv e-prints}\ ,\ \bibinfo {eid}
  {arXiv:2001.06683}} (\bibinfo {year} {2020})},\ \Eprint
  {https://arxiv.org/abs/2001.06683} {arXiv:2001.06683 [astro-ph.IM]}
  \BibitemShut {NoStop}%
\bibitem [{\citenamefont {{Astropy Collaboration}}\ \emph
  {et~al.}(2018)\citenamefont {{Astropy Collaboration}}, \citenamefont
  {{Price-Whelan}}, \citenamefont {{Sip{\H{o}}cz}}, \citenamefont
  {{G{\"u}nther}}, \citenamefont {{Lim}}, \citenamefont {{Crawford}},
  \citenamefont {{Conseil}}, \citenamefont {{Shupe}}, \citenamefont {{Craig}},
  \citenamefont {{Dencheva}}, \citenamefont {{Ginsburg}}, \citenamefont {{Vand
  erPlas}}, \citenamefont {{Bradley}}, \citenamefont {{P{\'e}rez-Su{\'a}rez}},
  \citenamefont {{de Val-Borro}}, \citenamefont {{Aldcroft}}, \citenamefont
  {{Cruz}}, \citenamefont {{Robitaille}}, \citenamefont {{Tollerud}},
  \citenamefont {{Ardelean}}, \citenamefont {{Babej}}, \citenamefont {others},\
  and\ \citenamefont {{Astropy Contributors}}}]{astropy}%
  \BibitemOpen
  \bibfield  {author} {\bibinfo {author} {\bibnamefont {{Astropy
  Collaboration}}}, \bibinfo {author} {\bibfnamefont {A.~M.}\ \bibnamefont
  {{Price-Whelan}}}, \bibinfo {author} {\bibfnamefont {B.~M.}\ \bibnamefont
  {{Sip{\H{o}}cz}}}, \bibinfo {author} {\bibfnamefont {H.~M.}\ \bibnamefont
  {{G{\"u}nther}}}, \bibinfo {author} {\bibfnamefont {P.~L.}\ \bibnamefont
  {{Lim}}}, \bibinfo {author} {\bibfnamefont {S.~M.}\ \bibnamefont
  {{Crawford}}}, \bibinfo {author} {\bibfnamefont {S.}~\bibnamefont
  {{Conseil}}}, \bibinfo {author} {\bibfnamefont {D.~L.}\ \bibnamefont
  {{Shupe}}}, \bibinfo {author} {\bibfnamefont {M.~W.}\ \bibnamefont
  {{Craig}}}, \bibinfo {author} {\bibfnamefont {N.}~\bibnamefont {{Dencheva}}},
  \bibinfo {author} {\bibfnamefont {A.}~\bibnamefont {{Ginsburg}}}, \bibinfo
  {author} {\bibfnamefont {J.~T.}\ \bibnamefont {{Vand erPlas}}}, \bibinfo
  {author} {\bibfnamefont {L.~D.}\ \bibnamefont {{Bradley}}}, \bibinfo {author}
  {\bibfnamefont {D.}~\bibnamefont {{P{\'e}rez-Su{\'a}rez}}}, \bibinfo {author}
  {\bibfnamefont {M.}~\bibnamefont {{de Val-Borro}}}, \bibinfo {author}
  {\bibfnamefont {T.~L.}\ \bibnamefont {{Aldcroft}}}, \bibinfo {author}
  {\bibfnamefont {K.~L.}\ \bibnamefont {{Cruz}}}, \bibinfo {author}
  {\bibfnamefont {T.~P.}\ \bibnamefont {{Robitaille}}}, \bibinfo {author}
  {\bibfnamefont {E.~J.}\ \bibnamefont {{Tollerud}}}, \bibinfo {author}
  {\bibfnamefont {C.}~\bibnamefont {{Ardelean}}}, \bibinfo {author}
  {\bibfnamefont {T.}~\bibnamefont {{Babej}}}, \bibinfo {author} {\bibnamefont
  {others}}, and\ \bibinfo {author} {\bibnamefont {{Astropy Contributors}}},\
  }\href {https://doi.org/10.3847/1538-3881/aabc4f} {\bibfield  {journal}
  {\bibinfo  {journal} {\aj}\ }\textbf {\bibinfo {volume} {156}},\ \bibinfo
  {eid} {123} (\bibinfo {year} {2018})},\ \Eprint
  {https://arxiv.org/abs/1801.02634} {arXiv:1801.02634 [astro-ph.IM]}
  \BibitemShut {NoStop}%
\bibitem [{\citenamefont {{Ginsburg}}\ \emph {et~al.}(2019)\citenamefont
  {{Ginsburg}}, \citenamefont {{Sip{\H{o}}cz}}, \citenamefont {{Brasseur}},
  \citenamefont {{Cowperthwaite}}, \citenamefont {{Craig}}, \citenamefont
  {{Deil}}, \citenamefont {{Guillochon}}, \citenamefont {{Guzman}},
  \citenamefont {{Liedtke}}, \citenamefont {{Lian Lim}}, \citenamefont
  {{Lockhart}}, \citenamefont {{Mommert}}, \citenamefont {{Morris}},
  \citenamefont {{Norman}}, \citenamefont {{Parikh}}, \citenamefont
  {{Persson}}, \citenamefont {{Robitaille}}, \citenamefont {{Segovia}},
  \citenamefont {{Singer}}, \citenamefont {{Tollerud}}, \citenamefont {{de
  Val-Borro}}, \citenamefont {{Valtchanov}}, \citenamefont {{Woillez}},
  \citenamefont {{Astroquery Collaboration}},\ and\ \citenamefont {{a subset of
  astropy Collaboration}}}]{astroquery}%
  \BibitemOpen
  \bibfield  {author} {\bibinfo {author} {\bibfnamefont {A.}~\bibnamefont
  {{Ginsburg}}}, \bibinfo {author} {\bibfnamefont {B.~M.}\ \bibnamefont
  {{Sip{\H{o}}cz}}}, \bibinfo {author} {\bibfnamefont {C.~E.}\ \bibnamefont
  {{Brasseur}}}, \bibinfo {author} {\bibfnamefont {P.~S.}\ \bibnamefont
  {{Cowperthwaite}}}, \bibinfo {author} {\bibfnamefont {M.~W.}\ \bibnamefont
  {{Craig}}}, \bibinfo {author} {\bibfnamefont {C.}~\bibnamefont {{Deil}}},
  \bibinfo {author} {\bibfnamefont {J.}~\bibnamefont {{Guillochon}}}, \bibinfo
  {author} {\bibfnamefont {G.}~\bibnamefont {{Guzman}}}, \bibinfo {author}
  {\bibfnamefont {S.}~\bibnamefont {{Liedtke}}}, \bibinfo {author}
  {\bibfnamefont {P.}~\bibnamefont {{Lian Lim}}}, \bibinfo {author}
  {\bibfnamefont {K.~E.}\ \bibnamefont {{Lockhart}}}, \bibinfo {author}
  {\bibfnamefont {M.}~\bibnamefont {{Mommert}}}, \bibinfo {author}
  {\bibfnamefont {B.~M.}\ \bibnamefont {{Morris}}}, \bibinfo {author}
  {\bibfnamefont {H.}~\bibnamefont {{Norman}}}, \bibinfo {author}
  {\bibfnamefont {M.}~\bibnamefont {{Parikh}}}, \bibinfo {author}
  {\bibfnamefont {M.~V.}\ \bibnamefont {{Persson}}}, \bibinfo {author}
  {\bibfnamefont {T.~P.}\ \bibnamefont {{Robitaille}}}, \bibinfo {author}
  {\bibfnamefont {J.-C.}\ \bibnamefont {{Segovia}}}, \bibinfo {author}
  {\bibfnamefont {L.~P.}\ \bibnamefont {{Singer}}}, \bibinfo {author}
  {\bibfnamefont {E.~J.}\ \bibnamefont {{Tollerud}}}, \bibinfo {author}
  {\bibfnamefont {M.}~\bibnamefont {{de Val-Borro}}}, \bibinfo {author}
  {\bibfnamefont {I.}~\bibnamefont {{Valtchanov}}}, \bibinfo {author}
  {\bibfnamefont {J.}~\bibnamefont {{Woillez}}}, \bibinfo {author}
  {\bibnamefont {{Astroquery Collaboration}}}, and\ \bibinfo {author}
  {\bibnamefont {{a subset of astropy Collaboration}}},\ }\href
  {https://doi.org/10.3847/1538-3881/aafc33} {\bibfield  {journal} {\bibinfo
  {journal} {\aj}\ }\textbf {\bibinfo {volume} {157}},\ \bibinfo {eid} {98}
  (\bibinfo {year} {2019})},\ \Eprint {https://arxiv.org/abs/1901.04520}
  {arXiv:1901.04520 [astro-ph.IM]} \BibitemShut {NoStop}%
\bibitem [{\citenamefont {{Foreman-Mackey}}\ \emph {et~al.}(2013)\citenamefont
  {{Foreman-Mackey}}, \citenamefont {{Conley}}, \citenamefont {{Meierjurgen
  Farr}}, \citenamefont {{Hogg}}, \citenamefont {{Lang}}, \citenamefont
  {{Marshall}}, \citenamefont {{Price-Whelan}}, \citenamefont {{Sanders}},\
  and\ \citenamefont {{Zuntz}}}]{Foreman-Mackey2013}%
  \BibitemOpen
  \bibfield  {author} {\bibinfo {author} {\bibfnamefont {D.}~\bibnamefont
  {{Foreman-Mackey}}}, \bibinfo {author} {\bibfnamefont {A.}~\bibnamefont
  {{Conley}}}, \bibinfo {author} {\bibfnamefont {W.}~\bibnamefont {{Meierjurgen
  Farr}}}, \bibinfo {author} {\bibfnamefont {D.~W.}\ \bibnamefont {{Hogg}}},
  \bibinfo {author} {\bibfnamefont {D.}~\bibnamefont {{Lang}}}, \bibinfo
  {author} {\bibfnamefont {P.}~\bibnamefont {{Marshall}}}, \bibinfo {author}
  {\bibfnamefont {A.}~\bibnamefont {{Price-Whelan}}}, \bibinfo {author}
  {\bibfnamefont {J.}~\bibnamefont {{Sanders}}}, and\ \bibinfo {author}
  {\bibfnamefont {J.}~\bibnamefont {{Zuntz}}},\ }\href@noop {} {\bibinfo
  {title} {{emcee: The MCMC Hammer}}} (\bibinfo {year} {2013}),\ \Eprint
  {https://arxiv.org/abs/1303.002} {ascl:1303.002} \BibitemShut {NoStop}%
\bibitem [{\citenamefont {{Hunter}}(2007)}]{matplotlib}%
  \BibitemOpen
  \bibfield  {author} {\bibinfo {author} {\bibfnamefont {J.~D.}\ \bibnamefont
  {{Hunter}}},\ }\href {https://doi.org/10.1109/MCSE.2007.55} {\bibfield
  {journal} {\bibinfo  {journal} {Computing in Science and Engineering}\
  }\textbf {\bibinfo {volume} {9}},\ \bibinfo {pages} {90} (\bibinfo {year}
  {2007})}\BibitemShut {NoStop}%
\bibitem [{\citenamefont {{van der Walt}}\ \emph {et~al.}(2011)\citenamefont
  {{van der Walt}}, \citenamefont {{Colbert}},\ and\ \citenamefont
  {{Varoquaux}}}]{numpy}%
  \BibitemOpen
  \bibfield  {author} {\bibinfo {author} {\bibfnamefont {S.}~\bibnamefont {{van
  der Walt}}}, \bibinfo {author} {\bibfnamefont {S.~C.}\ \bibnamefont
  {{Colbert}}}, and\ \bibinfo {author} {\bibfnamefont {G.}~\bibnamefont
  {{Varoquaux}}},\ }\href {https://doi.org/10.1109/MCSE.2011.37} {\bibfield
  {journal} {\bibinfo  {journal} {Computing in Science and Engineering}\
  }\textbf {\bibinfo {volume} {13}},\ \bibinfo {pages} {22} (\bibinfo {year}
  {2011})},\ \Eprint {https://arxiv.org/abs/1102.1523} {arXiv:1102.1523
  [cs.MS]} \BibitemShut {NoStop}%
\bibitem [{\citenamefont {{Virtanen}}\ \emph {et~al.}(2020)\citenamefont
  {{Virtanen}}, \citenamefont {{Gommers}}, \citenamefont {{Oliphant}},
  \citenamefont {{Haberland}}, \citenamefont {{Reddy}}, \citenamefont
  {{Cournapeau}}, \citenamefont {{Burovski}}, \citenamefont {{Peterson}},
  \citenamefont {{Weckesser}}, \citenamefont {{Bright}}, \citenamefont {{van
  der Walt}}, \citenamefont {{Brett}}, \citenamefont {{Wilson}}, \citenamefont
  {{Jarrod Millman}}, \citenamefont {{Mayorov}}, \citenamefont {{Nelson}},
  \citenamefont {{Jones}}, \citenamefont {{Kern}}, \citenamefont {{Larson}},
  \citenamefont {{Carey}}, \citenamefont {{Polat}}, \citenamefont {{Feng}},
  \citenamefont {{Moore}}, \citenamefont {{Vand erPlas}}, \citenamefont
  {{Laxalde}}, \citenamefont {{Perktold}}, \citenamefont {{Cimrman}},
  \citenamefont {{Henriksen}}, \citenamefont {{Quintero}}, \citenamefont
  {{Harris}}, \citenamefont {{Archibald}}, \citenamefont {{Ribeiro}},
  \citenamefont {{Pedregosa}}, \citenamefont {{van Mulbregt}},\ and\
  \citenamefont {{Contributors}}}]{scipy}%
  \BibitemOpen
  \bibfield  {author} {\bibinfo {author} {\bibfnamefont {P.}~\bibnamefont
  {{Virtanen}}}, \bibinfo {author} {\bibfnamefont {R.}~\bibnamefont
  {{Gommers}}}, \bibinfo {author} {\bibfnamefont {T.~E.}\ \bibnamefont
  {{Oliphant}}}, \bibinfo {author} {\bibfnamefont {M.}~\bibnamefont
  {{Haberland}}}, \bibinfo {author} {\bibfnamefont {T.}~\bibnamefont
  {{Reddy}}}, \bibinfo {author} {\bibfnamefont {D.}~\bibnamefont
  {{Cournapeau}}}, \bibinfo {author} {\bibfnamefont {E.}~\bibnamefont
  {{Burovski}}}, \bibinfo {author} {\bibfnamefont {P.}~\bibnamefont
  {{Peterson}}}, \bibinfo {author} {\bibfnamefont {W.}~\bibnamefont
  {{Weckesser}}}, \bibinfo {author} {\bibfnamefont {J.}~\bibnamefont
  {{Bright}}}, \bibinfo {author} {\bibfnamefont {S.~J.}\ \bibnamefont {{van der
  Walt}}}, \bibinfo {author} {\bibfnamefont {M.}~\bibnamefont {{Brett}}},
  \bibinfo {author} {\bibfnamefont {J.}~\bibnamefont {{Wilson}}}, \bibinfo
  {author} {\bibfnamefont {K.}~\bibnamefont {{Jarrod Millman}}}, \bibinfo
  {author} {\bibfnamefont {N.}~\bibnamefont {{Mayorov}}}, \bibinfo {author}
  {\bibfnamefont {A.~R.~J.}\ \bibnamefont {{Nelson}}}, \bibinfo {author}
  {\bibfnamefont {E.}~\bibnamefont {{Jones}}}, \bibinfo {author} {\bibfnamefont
  {R.}~\bibnamefont {{Kern}}}, \bibinfo {author} {\bibfnamefont
  {E.}~\bibnamefont {{Larson}}}, \bibinfo {author} {\bibfnamefont
  {C.}~\bibnamefont {{Carey}}}, \bibinfo {author} {\bibfnamefont
  {{\.I}.}~\bibnamefont {{Polat}}}, \bibinfo {author} {\bibfnamefont
  {Y.}~\bibnamefont {{Feng}}}, \bibinfo {author} {\bibfnamefont {E.~W.}\
  \bibnamefont {{Moore}}}, \bibinfo {author} {\bibfnamefont {J.}~\bibnamefont
  {{Vand erPlas}}}, \bibinfo {author} {\bibfnamefont {D.}~\bibnamefont
  {{Laxalde}}}, \bibinfo {author} {\bibfnamefont {J.}~\bibnamefont
  {{Perktold}}}, \bibinfo {author} {\bibfnamefont {R.}~\bibnamefont
  {{Cimrman}}}, \bibinfo {author} {\bibfnamefont {I.}~\bibnamefont
  {{Henriksen}}}, \bibinfo {author} {\bibfnamefont {E.~A.}\ \bibnamefont
  {{Quintero}}}, \bibinfo {author} {\bibfnamefont {C.~R.}\ \bibnamefont
  {{Harris}}}, \bibinfo {author} {\bibfnamefont {A.~M.}\ \bibnamefont
  {{Archibald}}}, \bibinfo {author} {\bibfnamefont {A.~H.}\ \bibnamefont
  {{Ribeiro}}}, \bibinfo {author} {\bibfnamefont {F.}~\bibnamefont
  {{Pedregosa}}}, \bibinfo {author} {\bibfnamefont {P.}~\bibnamefont {{van
  Mulbregt}}}, and\ \bibinfo {author} {\bibfnamefont {S.~.~.}\ \bibnamefont
  {{Contributors}}},\ }\href
  {https://doi.org/https://doi.org/10.1038/s41592-019-0686-2} {\bibfield
  {journal} {\bibinfo  {journal} {Nature Methods}\ }\textbf {\bibinfo {volume}
  {17}},\ \bibinfo {pages} {261} (\bibinfo {year} {2020})}\BibitemShut
  {NoStop}%
\end{thebibliography}%

\appendix
\section{Scaling Argument}
\label{app:scale}
In Section~\ref{OurFishingBoat}, we employ scaling arguments to show how the full survey sensitivity can be approximated by the MCMC results on a subset of stars in a single exposure. In this Appendix we derive these scaling relations explicitly. 

The log likelihood for a full dataset should be written as
\begin{equation}
    \ln \mathcal{L} = \sum _{i=1}^{N_m}\sum_{j=1}^{N_s}\sum_{k=0,1}\frac{(ds_{i,j,k}-dn_{i,j,k}(\Psi))^2}{2\sigma^2}\;,
\end{equation}
where the indices $i,j,k$ represents the exposures, the number of stars and the two components of the deflection vector. $ds$ is the data vector; in the null signal case, each data vector component follows a Gaussian distribution with zero mean and a standard deviation of $\sigma$. $dn(\Psi)$ is the predicted astrometric deflection given the GW source parameter $\Psi$. $\sigma$ is the telescope single-exposure single-source astrometric resolution. On average, 
\begin{equation}
\begin{split}
    \ln \mathcal{L} &\approx N_m N_s\sum_{k=0,1}\frac{\langle(ds_{k}-dn_{k}(\Psi))^2\rangle}{2\sigma^2}\\
    &= \sum_{k=0,1}\frac{\langle(ds_{k}-dn_{k}(\Psi))^2\rangle}{2(\sigma/ \sqrt{N_mN_s})^2}\;,
\end{split}
\end{equation}
where $\langle\cdot\rangle$ an average over the observed stars and the exposures. This is the same scaling relation in Section~\ref{AllFishingBoats} and Section~\ref{OurFishingBoat}, up to a constant scaling factor on the order of 1.

We note that the sensitivity is independent from the frequency of the GWs. By modeling the signal as purely sinusoidal, we numerically calculate the variance of the signal parameters using the Fisher information matrix. Once we have observed a significant number of signal cycles, the standard deviation of the wave amplitude approaches a constant, $\sqrt{2\pi/T}$, where $T$ is the total observation time. 
\end{document}